\documentclass[fleqn,usenatbib]{mnras}

\usepackage{newtxtext,newtxmath}

\usepackage[T1]{fontenc}

\DeclareRobustCommand{\VAN}[3]{#2}
\let\VANthebibliography\thebibliography
\def\thebibliography{\DeclareRobustCommand{\VAN}[3]{##3}\VANthebibliography}

\usepackage{graphicx}	
\usepackage{amsmath}	

\usepackage{subcaption}

\title[Variable pulsar emission and spin down]{The ubiquity of variable radio emission and spin-down rates in pulsars}

\author[M. E. Lower et al.]{\parbox{\textwidth}
{
M.~E. Lower,$^{1,2}$\thanks{E-mail: \href{mailto:mlower@swin.edu.au}{mlower@swin.edu.au}}
A.~Karastergiou,$^{3,4}$
S. Johnston$^{2}$,
P.~R. Brook,$^{5}$
S.~Dai,$^{2}$
M.~Kerr,$^{6}$
R.~N.~Manchester,$^{2}$\\
L.~S.~Oswald,$^{7}$
R.~M.~Shannon,$^{1,8}$
C.~Sobey,$^{9,10}$
and P.~Weltevrede$^{11}$
}
\\ \\
$^{1}$Centre for Astrophysics and Supercomputing, Swinburne University of Technology, PO Box 218, Hawthorn, VIC 3122, Australia\\
$^{2}$Australia Telescope National Facility, CSIRO, Space and Astronomy, PO Box 76, Epping, NSW 1710, Australia\\
$^{3}$Department of Astrophysics, University of Oxford, Denys Wilkinson Building, Keble Road, Oxford OX1 3RH, UK\\
$^{4}$Department of Physics and Electronics, Rhodes University, PO Box 94, Grahamstown 6140, South Africa\\
$^{5}$Institute for Gravitational Wave Astronomy and School of Physics and Astronomy, University of Birmingham, Edgbaston, Birmingham B15 2TT, UK\\
$^{6}$Space Science Division, Naval Research Laboratory, Washington, DC 20375, USA\\
$^{7}$School of Physics and Astronomy, University of Southampton, Southampton SO17 1BJ, UK\\
$^{8}$OzGrav: The ARC Centre of Excellence for Gravitational-wave Discovery\\
$^{9}$SKAO, ARRC Building, 26 Dick Perry Avenue, Kensington, WA 6151, Australia\\
$^{10}$CSIRO, Space and Astronomy, PO Box 1130 Bentley, WA 6102, Australia\\
$^{11}$Jodrell Bank Centre for Astrophysics, The University of Manchester, Alan Turing Building, Manchester, M13 9PL, United Kingdom
}

\date{Accepted XXXX. Received YYYY; in original form ZZZZ}

\begin{document}
\label{firstpage}
\pagerange{\pageref{firstpage}--\pageref{lastpage}}
\maketitle

\begin{abstract}
Pulsars are often lauded for their (relative) rotational and radio emission stability over long time scales. However, long-term observing programmes are identifying an increasing number of pulsars that deviate from this preconceived notion. Using Gaussian process regression and Bayesian inference techniques, we investigated the emission and rotational stability of 259 isolated radio pulsars that have been monitored using Murriyang, the Parkes 64\,m radio telescope, over the past three decades. We found that 238 pulsars display significant variability in their spin-down rates, 52 of which also exhibit changes in profile shape. Including 23 known state-switching pulsars, this represents the largest catalogue of variable pulsars identified to date and indicates that these behaviours are ubiquitous among the wider population. The intensity of spin-down fluctuations positively scales with increasing pulsar spin-down rate, with only a marginal dependence on spin-frequency. This may have substantial implications for ongoing searches for gravitational waves in the ensemble timing of millisecond pulsars. We also discuss challenges in explaining the physical origins of quasi-periodic and transient profile/spin-down variations detected among a subset of our pulsars. 
\end{abstract}

\begin{keywords}
radiation mechanisms: non-thermal -- stars: neutron -- pulsars: general
\end{keywords}


\section{Introduction}

The highly predictable, periodic nature of radio pulsars makes them excellent tools for studying a variety of astrophysical phenomena. 
These include measuring fluctuations in the electron column density and turbulence of the interstellar medium within our Galaxy \citep[e.g.][]{Petroff2013}, testing theories of gravity \citep{Freire2024}, and the direct detection of nanohertz-frequency gravitational waves \citep{Sazhin1978, Detweiler1979}.
The key to the success of such studies are the relative stability of both the pulsar emission mechanism and the rate at which pulsars spin down.
While individual pulses from a given pulsar can vary wildly in intensity and shape (e.g. \citealt{Johnston2001}), integrating over many rotations will typically return an average profile that is more or less constant from one observation to the next. 
Yet long-term monitoring programmes and observations with newer, more sensitive instruments have revealed a growing number of cases that deviate from ideal clock assumptions.

Mode or state switching between two or more discrete radio emission states was identified soon after the discovery of the first pulsars, manifesting as changes in the observed intensity, or shape or position in pulse longitude of the pulse profile components (see \citealt{Backer1970b, Backer1970c, Lyne1971}).
It is a broadband phenomenon, having been detected both at radio and X-ray wavelengths \citep{Hermsen2013, Hermsen2017, Hermsen2018}, and takes place over a wide range of timescales. 
This can be as short as one or several rotations of a neutron star, or slow secular variations that take place over many months or years. 
In some cases the emission appears to briefly switch off and back on again (nulls; \citealt{Backer1970a}) or remain off for extended periods of time (intermittency; \citealt{Kramer2006}).
Long-term monitoring experiments found that the timing of some pulsars displayed quasi-periodic processes with a range of apparent periods (e.g. \citealt{Cordes1981, Cordes1985, Stairs2000}).
Fluctuations in both radio emission and spin-down rate were eventually linked, initially through observations of the intermittent pulsar PSR~B1931$+$24 \citep{Kramer2006}, and later through increasingly large samples of pulsars that switch between two or more quasi-stable spin-down levels or undergo transient spin-down events 
\citep{Kramer2006, Hobbs2010, Lyne2010, Brook2014, Brook2016, Shaw2022}.
Other studies have revealed pulsars with variations in either spin-down or pulse shape (and even polarization) that appear uncorrelated with one another \citep{Brook2016, Shaw2022, Basu2024}, though this may be a selection effect imparted by limited telescope sensitivity and observing cadence.

Exactly what mechanism drives these behaviours, and whether they originate from the interior dynamics or magnetospheric fluctuations of neutron stars, is presently unknown.
Free precession of a non-axisymmetric neutron star was once a popular hypothesis for explaining highly periodic changes in pulsar profile shapes and spin-down rates \citep{Stairs2000}.
However, the discovery that such spin-down rate changes were linked to short-term emission variations has cast doubt on whether free precession is the true, or only, mechanism behind these behaviours \citep{Lyne2010, Jones2012, Kerr2016, Stairs2019}.
Simultaneous emission switching at radio and X-ray wavelengths point toward a magnetospheric reconfiguration being responsible for at least some of the mode-switching behaviour \citep{Hermsen2013}. 
Emission and spin-down variations that seemingly coincided with glitches in several pulsars (e.g, \citealt{Weltevrede2011, Keith2013}) may point to crustal events such as starquakes that couple the internal dynamics of the neutron star to the magnetosphere \citep{Akbal2015, Yuan2021}. 
Indeed, the 2016 glitch in the Vela pulsar (PSR~J0835$-$4510/B0833$-$45) was associated with a short-lived quenching of its radio emission \citep{Palfreyman2018}, speculated to be due to a quake-induced magnetic disturbance \citep{Bransgrove2020}.
External triggers such as interactions between pulsars and minor bodies are another popular hypothesis.
These include interactions with individual in-falling asteroids \citep{Brook2014}, debris disks \citep{Li2006, Shannon2013, Jennings2020b}, and impacts from interstellar objects \citep{Pham2024}.
In such cases, the ionised remains of a wayward minor body alter the plasma content of the pulsar magnetosphere, subsequently enhancing or attenuating the radio emission and altering the spin-down state.
Many previous attempts to explain pulsar variability focussed on either individual objects or small samples with similar characteristics \citep[e.g.,][]{Brook2014, Brook2016, Lower2023, Basu2024}.
Understanding whether one or more of these processes are responsible for these profile/spin-down variations requires population-level studies of a large ensemble of pulsars.

Determining the prevalence of pulsar rotation/emission state changes across the population is vital to building a complete picture of the processes that contribute to the overall noise budget of a pulsar.
Knowing how these behaviours scale as functions of spin frequency and spin-down rate is particularly important for pulsar timing array experiments that utilise an ensemble of millisecond pulsars to search for nanohertz-frequency gravitational waves \citep{Manchester2013a, Manchester2013b, Kramer2013, McLaughlin2013}. 
Even the most stable millisecond pulsars have been found to display profile shape variations and emission state switching that can hamper such experiments \citep{Shannon2016, Brook2018, Goncharov2021, Miles2022, Nathan2023, Jennings2024}.
It may also bring us a step closer to determining the origins of unexpected time-dependent changes in the common noise process seen by pulsar timing array experiments (see \citealt{Reardon2023}).

In this work, we present the largest sample of pulsars found to display emission and spin-down variations to date, by applying the Gaussian process regression techniques of \citep{Brook2016, Brook2018} to observations collected over the past three decades by Murriyang, the CSIRO Parkes 64 m radio telescope.
In Section~\ref{sec:obs} we briefly describe the data collection and reduction process.
Section~\ref{sec:methods} details the methodologies used for obtaining updated pulsar timing solutions and modelling the pulse profile shape and spin-down variations, the results of which are presented in Section~\ref{sec:results}. 
We discuss how these effects vary across the population and the implications for pulsar timing array experiments in Section~\ref{sec:disc}.
Finally, we summarise the results and make concluding statements in Section~\ref{sec:conc}.

\section{Observations and data reduction}\label{sec:obs}

\begin{figure}
    \centering
    \includegraphics[width=\linewidth]{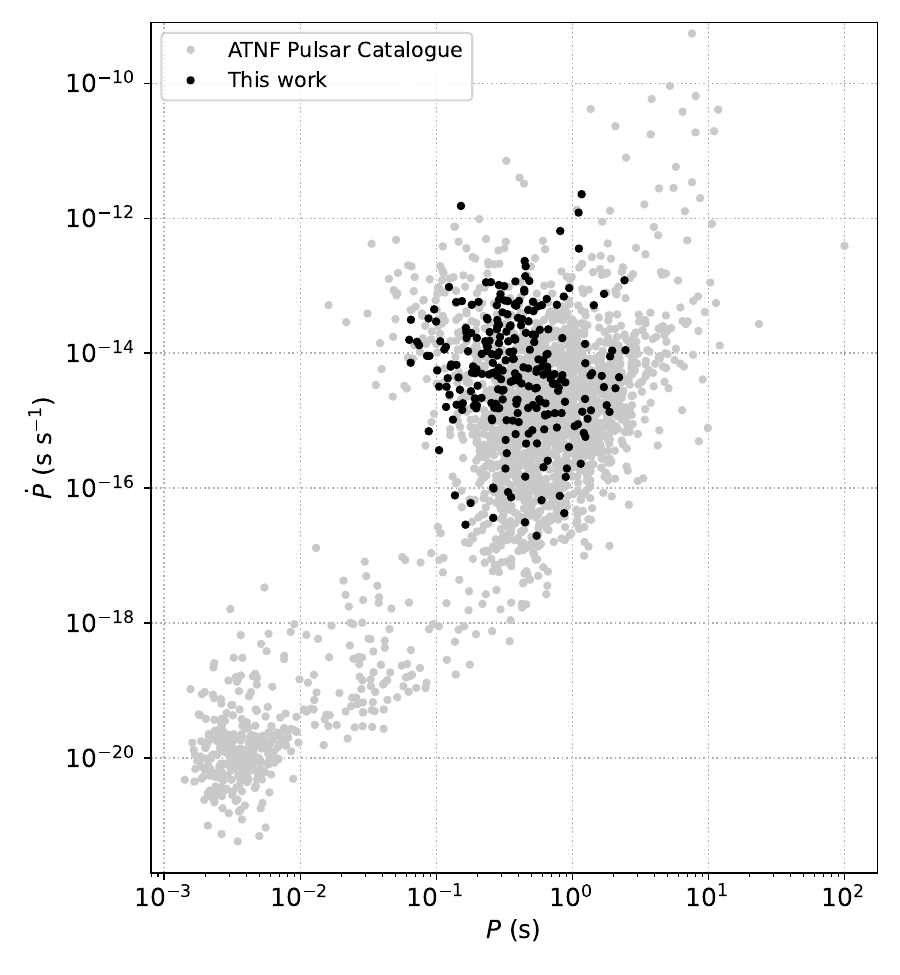}
    \vspace{-0.5cm}
    \caption{A pulsar period ($P$) and period-derivative ($\dot{P}$) diagram, where the pulsars sample analysed in this work are highlighted by black circles. Light-grey circles indicate all known pulsars published in v2.5.1 of the ATNF Pulsar Catalogue.}
    \label{fig:ppdot}
\end{figure}

The pulsar data analysed in this work were collected using Murriyang between 2007-2023 under the `young' pulsar timing (P574) programme \citep{Weltevrede2010, Johnston2021a}.
Our sample covers 259 isolated, non-recycled, rotation-powered pulsars with characteristic ages ranging from 1.6\,kyr (PSR~J1513$-$5908) to 0.4\,Gyr (PSR~J0820$-$4114).
In Figure~\ref{fig:ppdot} we show the positions of our pulsar sample in period-period-derivative space alongside those of the broader ATNF Pulsar Catalogue \citep{Manchester2005}\footnote{\href{https://www.atnf.csiro.au/research/pulsar/psrcat/}{https://www.atnf.csiro.au/research/pulsar/psrcat/}}.
Note that there is a slight bias towards pulsars with high rotational kinetic energy loss rates due to the programme's continued support of the Fermi mission \citep{Smith2023}. 
Pulsar observations were undertaken with the 10/50\footnote{Later re-tuned to cover the 40-cm band due to increased interference.}, 20-cm Multibeam, H-OH and Ultra-Wideband Low (UWL) receivers \citep{Granet2001, Staveley-Smith1996, Granet2011, Hobbs2020} and saved to {\sc psrfits} format archives \citep{Hotan2004, vanStraten2012}.
Details of the specific observing set-up, radio-frequency excision, and calibration steps can be found in \citet{Namkham2019, Parthasarathy2019, Lower2021b, Johnston2021a}.
Where possible, we also made use of extended timing datasets collected using legacy signal processors going back to the 1990s, which are available in the Parkes Observatory Pulsar Data Archive \citep{Hobbs2011}\footnote{\href{https://data.csiro.au/domain/atnf}{https://data.csiro.au/domain/atnf}}. 

Pulse times of arrival (ToAs) were obtained following the prescription outlined previously in \citet{Parthasarathy2019}, \citet{Lower2021b} and \citet{Lower2023}.
We averaged the individual observations in time and frequency to form one-dimensional pulse profiles, which were then cross-correlated against a noise-free template using the Fourier-domain Monte Carlo (FDM) method built into the {\sc pat} tool in {\sc psrchive} \citep{Hotan2004, vanStraten2012}. 
Note that we employed observing band-specific templates for each pulsar.

\section{Methods}\label{sec:methods}

\subsection{Updated pulsar timing}

A substantial fraction of the pulsars required updates to their timing models over those published in the ATNF Pulsar Catalogue.
Phase coherent solutions were derived following the manual procedures outlined in \citet{Parthasarathy2019} and \citet{Lower2021b}, where pulsar positions, proper motions, spin frequency ($\nu$), spin-down rate ($\dot{\nu}$) and on occasion, second spin-frequency derivative ($\ddot{\nu}$), were re-fit using a combination of {\sc tempo2} and {\sc temponest} \citep{Hobbs2006, Lentati2014}.
The majority of pulsars that required inclusion of $\ddot{\nu}$ were those previously studied in \citet{Parthasarathy2020} and \citet{Lower2021b}.
Glitch events were identified by eye as discontinuities or sharp features in the timing residuals.
These were corrected for by fitting the corresponding changes in $\nu$ and $\dot{\nu}$ using {\sc tempo2}.
Obtaining accurate measurements of the recovery parameters is challenging and imperfectly removed recovery signals will introduce unwanted artefacts in our spin-down analysis.
A detailed analysis of the new glitch events and overall updates to the timing of these pulsars is beyond the scope of this work, and will be reported elsewhere.
Once a coherent solution was obtained, we then applied pulse numbering to the ToAs to maintain an accurate record of every single rotation over our timing baseline.
In Figure~\ref{fig:resids} we show the corresponding timing residuals for our final sample of 259 pulsars.

\begin{figure*}
    \centering
    \includegraphics[width=\linewidth]{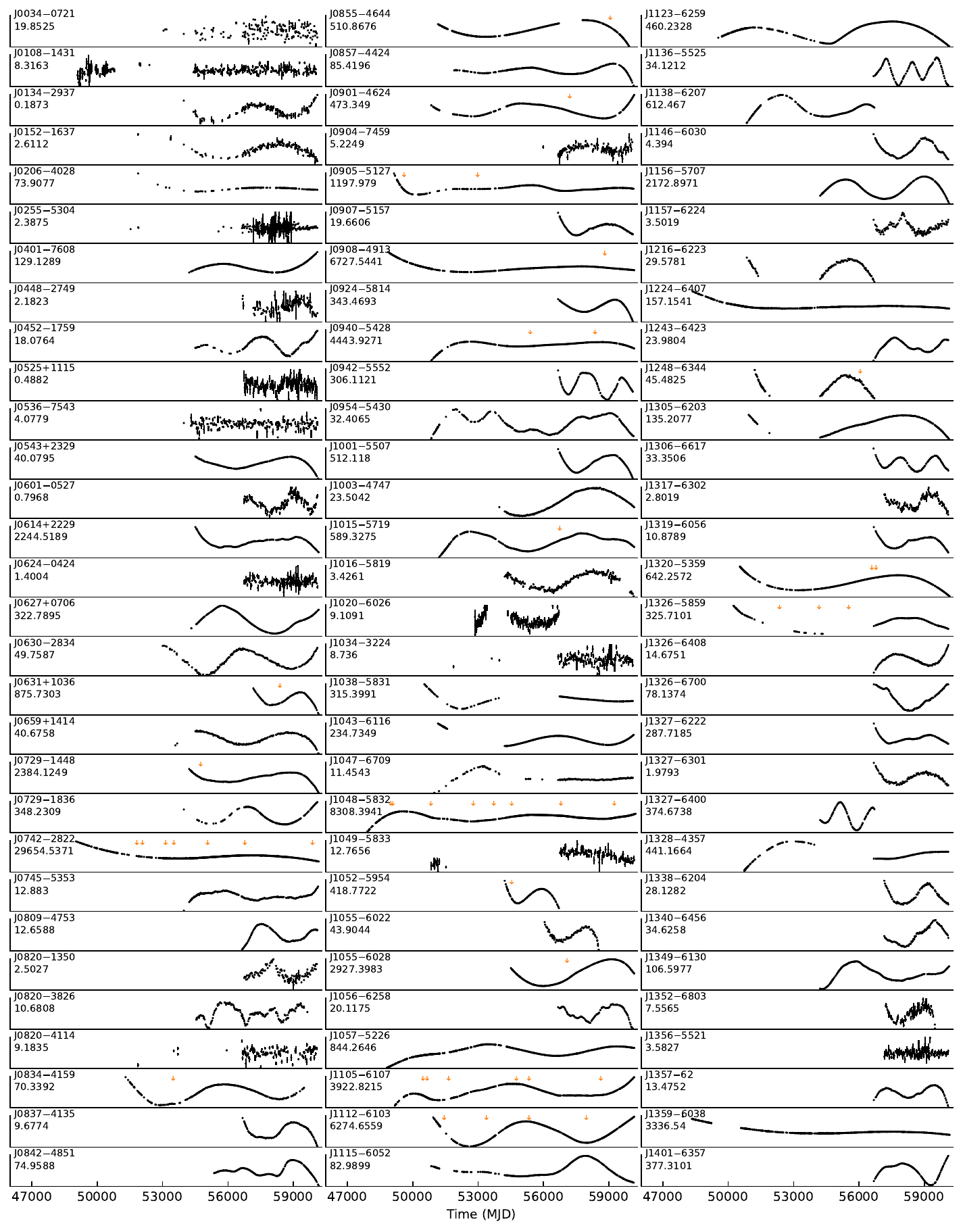}
    \vspace{-0.5cm}
    \caption{Timing residuals for all 259 pulsars after subtracting the best-fit timing model. Labels on the left-hand side of each panel indicates the pulsar J2000 name and the maximum-to-minimum range of the residuals in milliseconds. The downward pointing arrows indicate the epochs of detected glitches.}
    \label{fig:resids}
\end{figure*}

\begin{figure*}
    \centering
    \includegraphics[width=\linewidth]{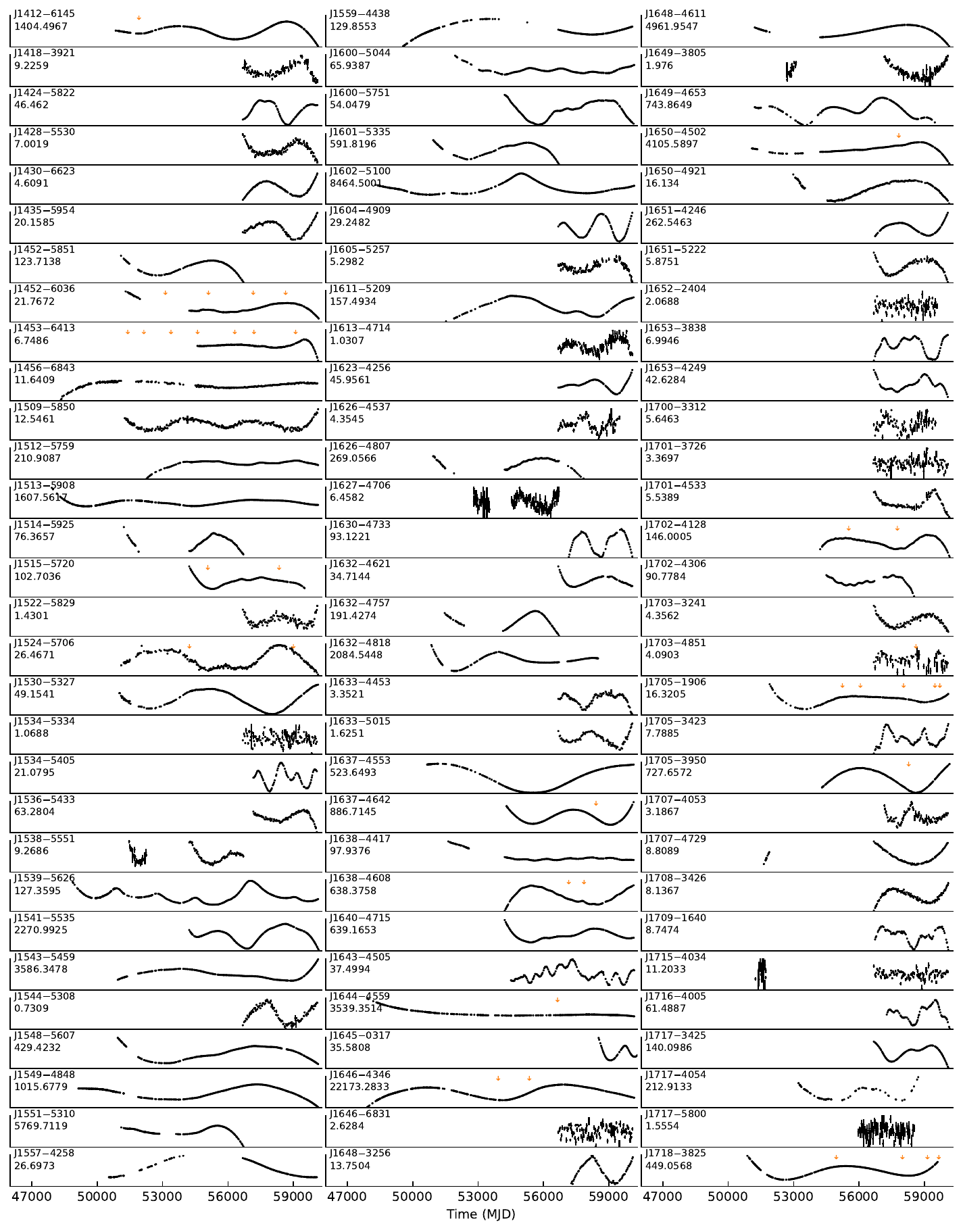}
    \vspace{-0.5cm}
    \contcaption{}
\end{figure*}

\begin{figure*}
    \centering
    \includegraphics[width=\linewidth]{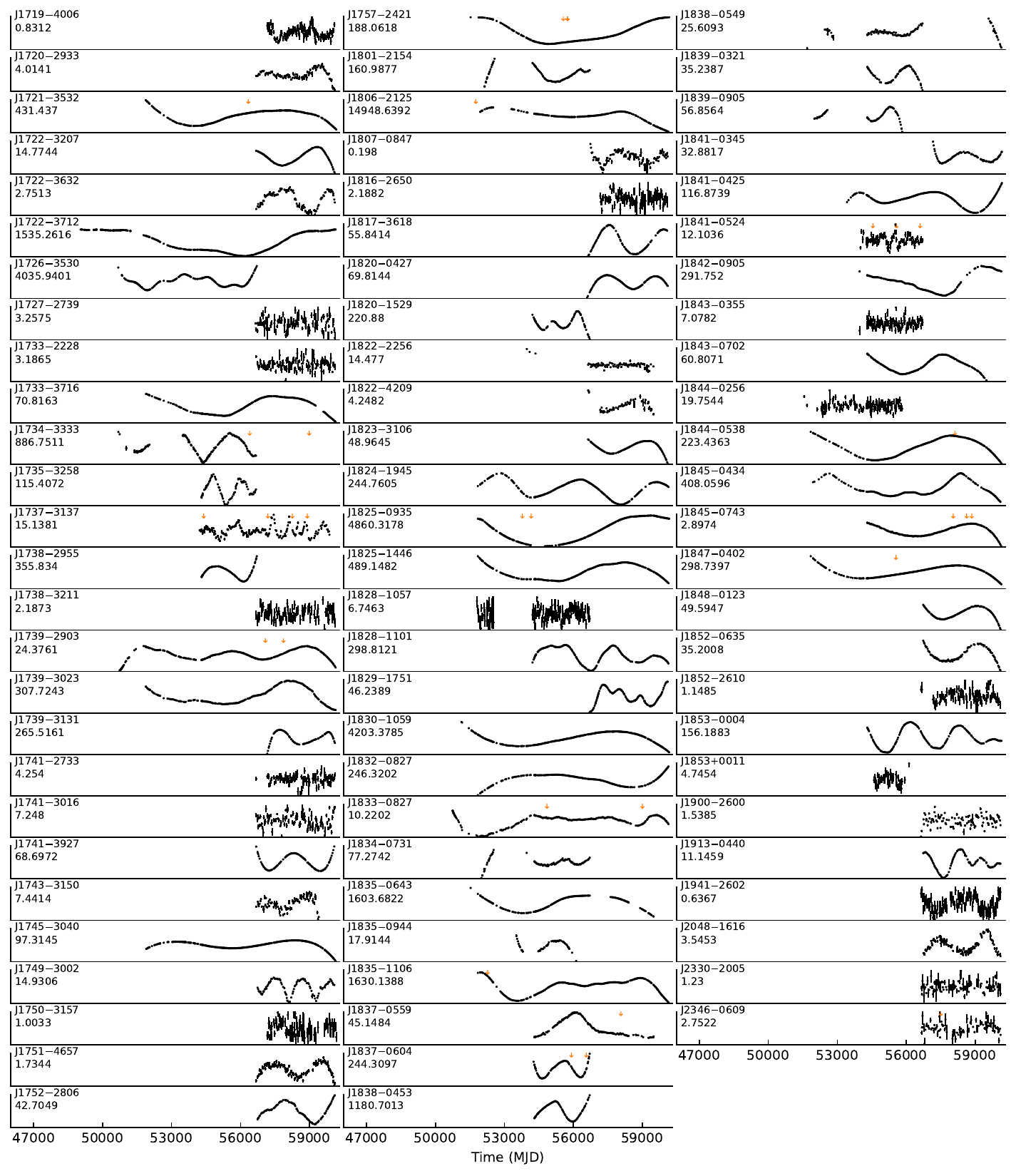}
    \vspace{-0.5cm}
    \contcaption{}
\end{figure*}

\subsection{Profile variability maps}

In order to model the observed variations in the pulsar profiles and spin-down rates, we followed the general methodology devised by \citet{Brook2016}.
To detect and then study changes in the radio profile shape, we created profile variability maps that display variations in profile intensity as functions of rotational phase and observation date.
Only the total intensity radio profiles collected by the Parkes Multibeam, H-OH and 1216--1600\,MHz UWL subband were used in the shape variation analysis.
Observations that were visually affected by residual RFI, instrumental artefacts, or displayed an anomalously low signal-to-noise (S/N) ratio were manually flagged and excluded.
As in \citet{Lower2023}, we performed an initial coarse alignment of the profiles by maximising the correlation between each observation with the highest S/N profile.
The alignment was then refined using the iterative scaling and shifting approach outlined in \citet{Brook2018}.
We created standard templates for each pulsar by computing the median normalised flux within each phase bin, which were then subtracted from our observations to produce profile residuals.
These profile residuals were then fitted with non-parametric models on a per phase bin basis via Gaussian process regression to produce profile variability maps.
Variability maps where the absolute peak phase-bin standard deviation ($\max|\sigma_{\rm prof}|$) was greater than three times the median off-pulse root-mean-square often indicated the presence of substantial profile shape variability above random noise fluctuations in the data.
However, this metric is sensitive to distortions in the data caused by residual RFI, hence visual inspection of the variability maps and profiles was used to filter out these outliers.
We also note that apparent shape changes due to the different central frequencies of the multibeam (2007-2016, 2017-2019), H-OH (2016-2017) and UWL (2019 onwards) receivers limits our sensitivity to more subtle temporal shape changes.

We made use of the Mat\'{e}rn and white noise kernels implemented in {\sc scikit-learn} \citep{scikit-learn} for modelling the differences in profile shapes after subtracting a median template.
This is the same kernel that was used by \citet{Shaw2022}, which is better at capturing short-duration profile shape changes than the squared-exponential kernel.
For the Mat\'{e}rn kernel, we used a positive covariance exponent of $\mu = 3/2$
\begin{equation}
    C_{3/2}(t, t') = \sigma_{f}^{2} \Bigg( 1 + \frac{\sqrt{3}\,|t - t'|}{\lambda} \Bigg) \exp \Bigg( -\frac{\sqrt{3}\,|t - t'|}{\lambda} \Bigg),
\end{equation}
where $\sigma_{f}^{2}$ is the maximum allowed covariance, $|t - t'|$ is the time separating two data points, and $\lambda$ is the kernel length scale.
We computed the optimal kernel hyperparameters using a maximum-likelihood approach where the kernel length scale was allowed to vary between $30\,{\rm d} < \lambda < 300\,{\rm d}$. 
An exception is PSR~J1359$-$6038, which required the use of two separate kernels with respective length scales of $30\,{\rm d} < \lambda_{1} < 600\,{\rm d}$ and $600\,{\rm d} < \lambda_{2} < 3000\,{\rm d}$ to correctly capture the short and long-term profile variations displayed by this pulsar.
In general, we found the {\sc scikit-learn} implementations of the Mat\'{e}rn and white noise kernels provided more accurate representations of the underlying profile variations in our pulsars when compared to a similar variant implemented in the {\sc george} Python package \citep{Ambikasaran2015}.
The latter would often miss rapid changes in the profile shape. 

\subsection{Modelling spin-down rate variations}

Our approach to modelling the spin-down rates followed that of \citet{Shaw2022}, where we used the {\sc gp\_nudot.py} script in {\sc PulsarPVC}\footnote{\href{https://gitlab.com/benjaminshaw/pulsarpvc}{https://gitlab.com/benjaminshaw/pulsarpvc}} to fit a squared exponential kernel to the timing residuals for each pulsar.
Spin-down timeseries were then generated by taking the second derivative of the resulting model of the timing residuals.
As with the profile variability, this made use of the Gaussian process regression functionality of {\sc scikit-learn}.
Similar to the analyses of \citet{Brook2016} and \citet{Shaw2022}, about half of our pulsars required the addition of a second squared exponential kernel to account for both long- and short-term variations. 
Several dozen pulsars in our sample displayed a sinusoidal signal with a 1-yr period in their resulting spin-down timeseries that resulted from small offsets from their assumed position on the sky.
For some pulsars, this was corrected through updated timing model parameters obtained via {\sc tempo2} and {\sc TempoNest}.
Yet for others, a degeneracy between the annual sinusoid and quasi-periodic pulsar timing noise, combined with the limitations of the red power-law timing noise model in {\sc TempoNest}\footnote{Modelling of quasi-periodic and non-stationary timing behaviours as a red power-law process can result in biased measurements of the pulsar properties (cf. \citealt{Keith2023}).}, prevented us from making such positional corrections for a substantial number of pulsars. 
To overcome this, we also conducted a set of Gaussian process regression fits that included a sinusoidal kernel with a fixed 1-yr period.
In order to determine which pulsars had residuals best described by a one-kernel or two-kernel fit, or required an annual sinusoid to correct for positional offsets, we compared the Bayesian information criterion (BIC) for each model.
We first computed the BIC as
\begin{equation}
    {\rm BIC} = k \ln(N)-2\ln(\mathcal{L}_{\max}),
\end{equation}
where $k$ is the number of free parameters in the model, $N$ is the number of data points, and $\mathcal{L}_{\max}$ is the maximum likelihood value for the data given the model.
A smaller BIC indicates a better match between the model and the data.
The statistical significance for which one model is preferred over the other can be inferred from the difference in BICs for each model
\begin{equation}
    \Delta{\rm BIC} = {\rm BIC}(\mathcal{M}_{\rm 2})-{\rm BIC}(\mathcal{M}_{\rm 1}),
\end{equation}
where the subscripts ${\rm 2}$ and ${\rm 1}$ refer to the two models being compared.
The second model is preferred when $\Delta{\rm BIC} < 0$, while $\Delta{\rm BIC} > 0$ favours the first model.
This BIC-informed model selection approach worked well for most pulsars.
However, on occasion we had to make by-eye judgement calls when one model visually matched the data better than another in spite of the reported BIC.
This often occurred in instances where two squared exponential kernels and a sinusoidal kernel were needed to fit both the spin-down variations plus an incorrect position. 
After determining which set of kernels provided the best match to the timing residuals, continuous models of the pulsar spin-down rate were computed by taking the second derivative of the resulting non-parametric model (see \citealt{Brook2016} for details).

Identifying spin-down variations through visual inspection of the $\dot{\nu}$ timeseries is relatively straightforward.
However, to avoid biases in instances where the fluctuation amplitude is comparable to the derived uncertainties we computed a significance metric for each pulsar.
This `$\mathcal{K}$-metric` is given by
\begin{equation}
    \mathcal{K} = \frac{|\dot{\nu}_{\rm min}| - |\dot{\nu}_{\rm max}|}{2\sigma_{\nu, {\rm mean}}}
\end{equation}
where $\dot{\nu}_{\rm min}$ and $\dot{\nu}_{\rm max}$ refer to the minimum and maximum inferred absolute spin-down rates derived from the Gaussian process regression, and $\sigma_{\dot{\nu}, {\rm mean}}$ is the mean spin-down uncertainty computed via equations 9 and 10 of \citet{Brook2016}.
We used a threshold of $\mathcal{K} > 1$ for defining when a pulsar displayed substantial spin-down variability.
To test the level of correlation between any observed variations in profile shape and spin-down rate, we computed the Spearman rank correlation coefficient between each phase bin of the profile variability map and $\dot{\nu}$ for time-lags ranging between $\pm$ half the length of the per-pulsar observing span \citep{Brook2016, Shaw2022, Lower2023}.

\section{Results and analysis}\label{sec:results}

Our variability analysis identified 238 pulsars that displayed significant ($\mathcal{K} > 1$) changes in $\dot{\nu}$ over time.
We present the resulting spin-down timeseries for the 238 variable $\dot{\nu}$ pulsars in Figure~\ref{fig:nudot}, and list the recovered Gaussian process hyper-parameters and inferred rotational properties in Tables~\ref{tbl:nudot_fit} and \ref{tbl:magnetosphere} in Appendix~\ref{appndx}.
In general, we find the largest changes in spin-down can be broadly described by four different categories: 
\begin{enumerate}
    \item quasi-periodic $\dot{\nu}$ variations with one or multiple characteristic period(s),
    \item a constantly changing value of $\dot{\nu}$ that does not display clear quasi-periodic behaviour,
    \item long-term smooth variations in $\dot{\nu}$ over long timescales and
    \item transient increases or decreases in $\dot{\nu}$.
\end{enumerate}

Profile variability maps were generated for 214 pulsars that displayed significant $\dot{\nu}$ variations, which can be found in the Supplementary Materials.
One such example is shown in Figure~\ref{fig:var_J1048} for PSR~J1048$-$5832, where the upper and lower panels present the varying spin spin-down rate and profile shape changes respectively.
The remaining 24 of the 238 variable pulsars had insufficient per-epoch $S/N$ to generate accurate variability maps.
From these maps, we found that 52 pulsars showed substantial changes in profile shape.
We also identified several pulsars that appear to show intense epoch-to-epoch shape variations across the entire profile that are approximately symmetric about the peak.
This behaviour can be attributed to profile jitter, resulting from the finite number of rotations that are recorded in our relatively short, per-pulsar observations. 
Pulsars that were most strongly affected by this include PSRs J0837$-$4135, J1430-6623 and J1644-4559, which are not included in the profile shape change analysis.

\begin{figure*}
    \centering
    \includegraphics[width=\linewidth]{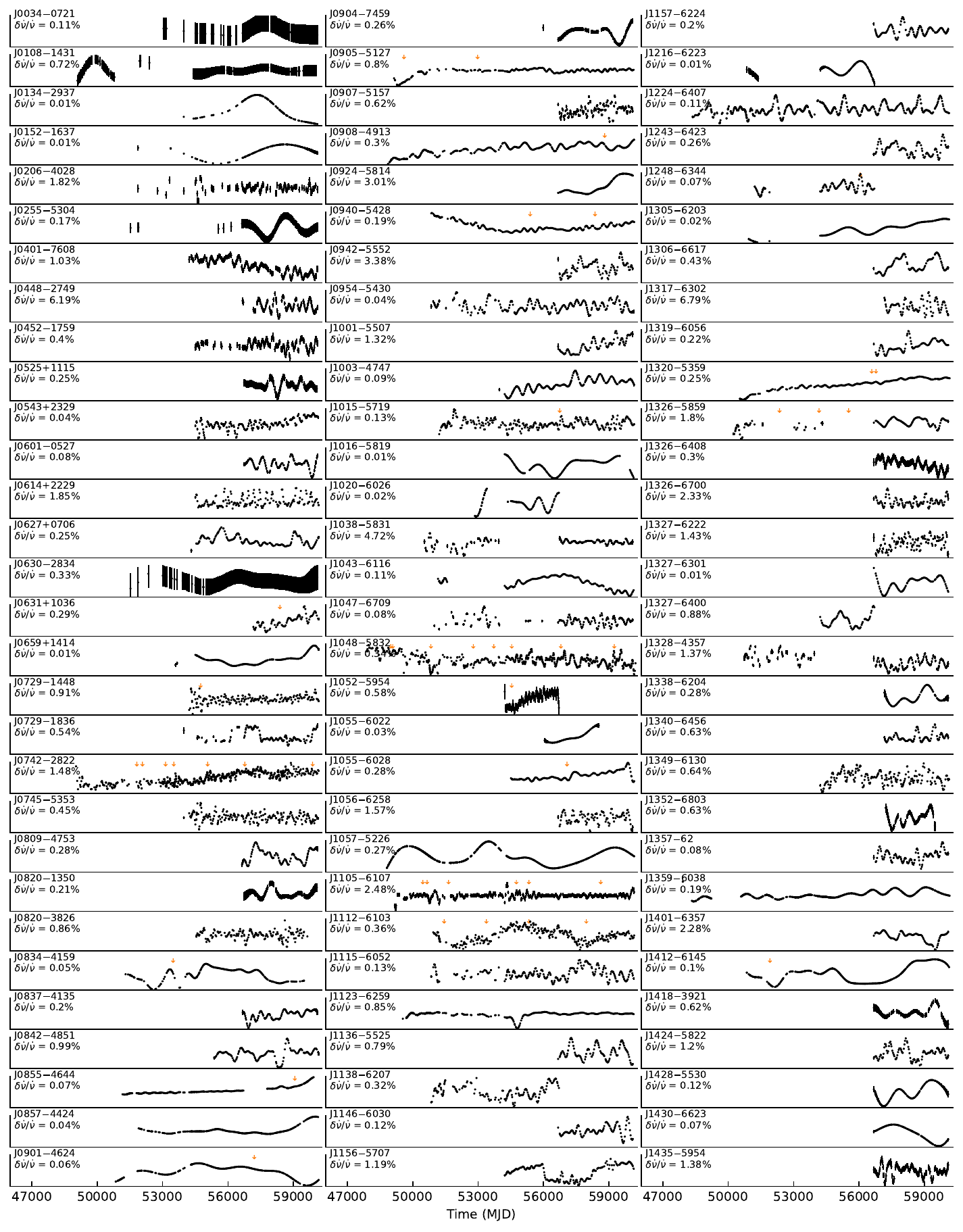}
    \vspace{-0.5cm}
    \caption{Spin-down timeseries for the 238 pulsars that displayed significant variability. Labels on the left-hand side of each panel indicates the pulsar J2000 name and the percentage difference between the minimum and maximum value of $\dot{\nu}$. The downward pointing arrows indicate the epochs of detected glitches.}
    \label{fig:nudot}
\end{figure*}

\begin{figure*}
    \centering
    \includegraphics[width=\linewidth]{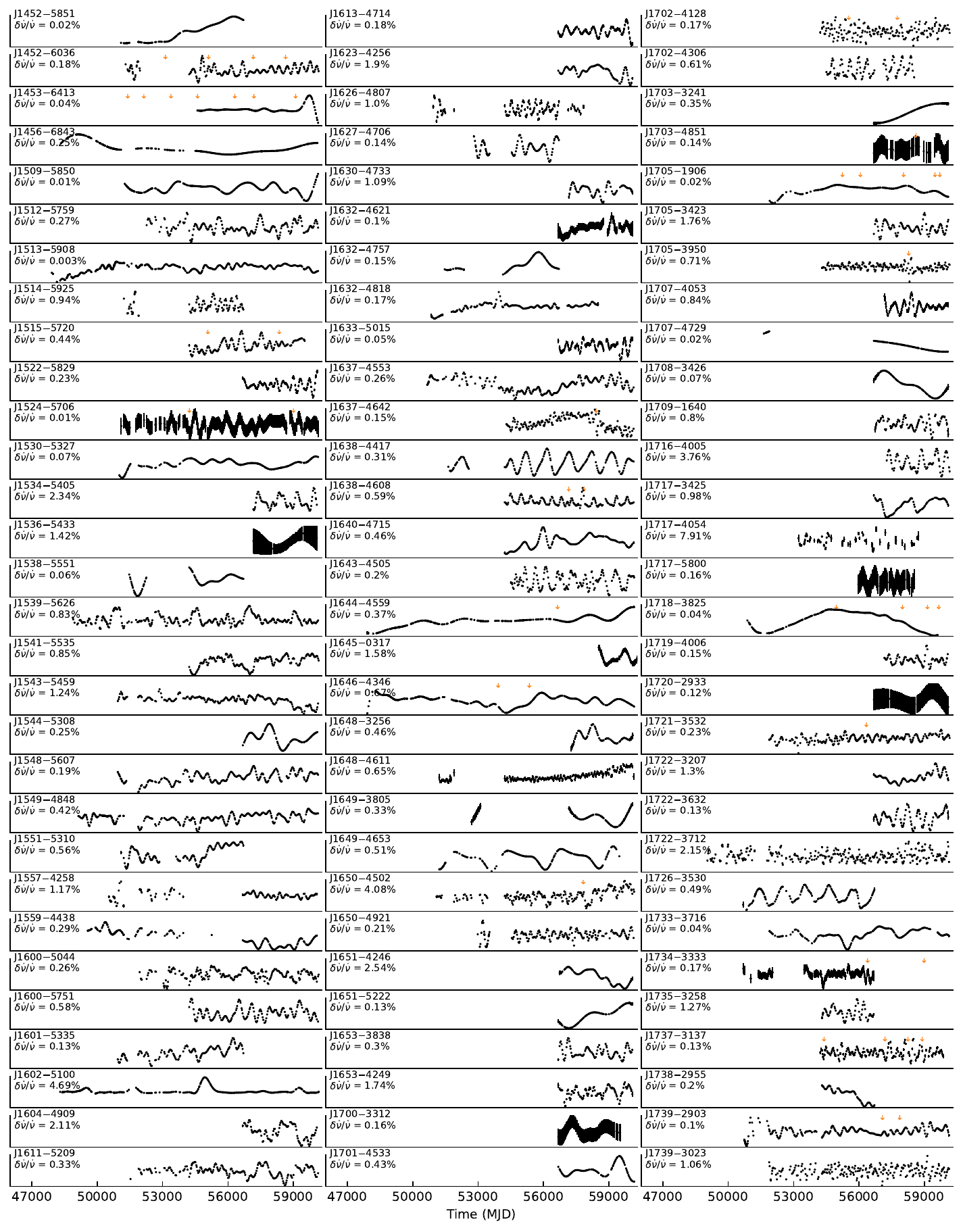}
    \vspace{-0.5cm}
    \contcaption{}
\end{figure*}

\begin{figure*}
    \centering
    \includegraphics[width=\linewidth]{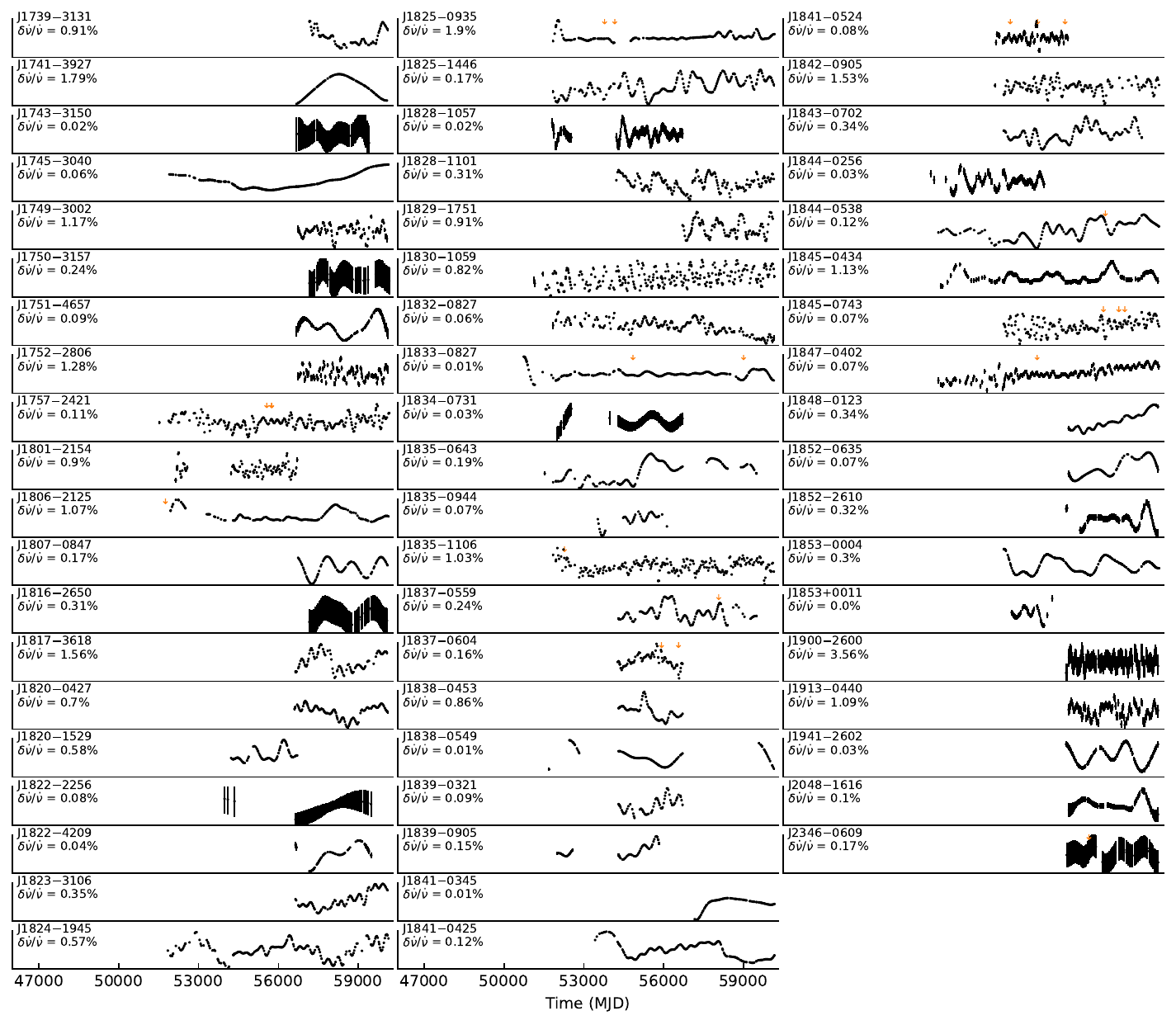}
    \vspace{-0.5cm}
    \contcaption{}
\end{figure*}

\subsection{Known variable pulsars}

Among our sample are 23 pulsars that had been previously identified as displaying correlated profile and spin-down variations, or emission state switching in single-pulse observations.
Here we provide brief summaries of how these pulsars have either continued to vary with time, or display newly discovered variability.

\subsubsection*{PSR J0742$-$2822 (B0740$-$28)}

Among the initial sample of variable profile/spin-down pulsars reported by \citet{Lyne2010}, PSR J0742$-$2822 has a fast variability timescale of only $\sim 135$\,d.
The relationship between changes in profile shape and spin-down rate of this pulsar is complicated. 
Both are largely uncorrelated on long-time scales, but display an increased levels of correlation following the MJD 55022 glitch (see, \citealt{Keith2013}), suggested to be a result of glitch-induced coupling between the internal and external dynamics of the pulsar.
This behaviour is readily apparent in the variability map as deficits (excess) emission from the leading (dip between leading and trailing) profile components.
The increased correlation appears to vanish $\sim$330\,d before the MJD 56727 glitch, and fails to re-appear in the post-glitch spin-down/shape variations. 
Similarly, the MJD 59839 glitch also does not result in enhanced correlation between the profile and spin-down rate, indicating the phenomenon identified by \citet{Keith2013} is not a universal feature of the post-glitch behaviour of PSR~J0742$-$2822.

\subsubsection*{PSR J0729$-$1448}

The main peak in this pulsar's profile was found to have switched from a brighter emission state to a weak state in MeerKAT observations taken around MJD~59400 by \citet{Basu2024}.
No clear correlation was identified between this state change and variations in the pulsar spin-down rate.
Our less sensitive Murriyang observations do not reveal any noticeable profile shape variations, however the spin-down modelling recovers rapid $0.91\%$ fluctuations in $\dot{\nu}$. 
A Lomb-Scargle periodogram generated from the spin-down timeseries recovered an apparent quasi-period of $\sim$183\,d. 

\subsubsection*{PSR J0614$+$2229 (B0611$+$22)}

This 0.33\,s pulsar that has been reported to switch between two emission states: `mode A' where the emission occurs slightly earlier in pulse phase, and `mode B' where it is both weaker and arrives slightly later \citep{Zhang2020}.
Our spin-down model displays rapid, short-duration decreases in spin-down rate with a quasi-periodic spacing of $\sim$268\,d.
These spin-down events typically last between 130-200\,d with typical amplitudes of $\delta\dot{\nu}/|\dot{\nu}| \sim 1.58$\%.
The profile variability map shows the trailing edge of the profile displays changes on a similarly rapid timescale, usually appearing as an excess of emission whenever the pulsar is in the lower spin-down state that is only mildly correlated with the spin-down timeseries.
An excess of emission in this part of the profile would point to the pulsar preferentially emitting in `mode B' between the quasi-periodic decreases in spin-down rate.

\subsubsection*{PSR J0908$-$4913 (B0906$-$49)}

Rocketing away from the putative supernova remnant G270.4$-$1.0 \citep{Johnston2021b}, this orthogonally rotating pulsar was found by \citet{Brook2016} to display correlated changes in the leading edge of the main pulse and both inter-pulse components. 
All three of these profile components display simultaneous excesses and deficits of emission, which are also reflected in our updated profile variability map.
Our profile/spin-down correlation map shows a positive correlation between the changes in these profile components that lag corresponding spin-down variations by $\sim$300\,d, while the more rapid changes in the peak of the main-pulse being uncorrelated with spin down rate.
Similar cross-talk between emission originating from antipodal magnetic poles was previously identified in the single-pulse mode switching of PSRs J1057$-$5226, J1705$-$1906 and J1825$-$0935\footnote{Note that model fits to the linear polarisation position angle swing of PSR~J1825$-$0935 are consistent with both an aligned or orthogonal rotator \citep{Johnston2023}.} \citep{Weltevrede2007, Weltevrede2012, Fowler1982}, indicating emission state-switching is a global phenomenon that affects the plasma content of the entire magnetosphere.
We also detect more rapid variations in the primary peak of the main-pulse, though this may be the result of profile jitter shifting the exact peak of the profile from epoch-to-epoch.
A small glitch occurred in PSR J0908$-$4913 on MJD 58765 \citep{Lower2021b}, which does not appear to have had a noticeable impact on the post-glitch profile shape or spin-down variability.

\subsubsection*{PSR J0940$-$5428}

This pulsar was noted by \citet{Brook2016} as an example where changes in spin-down over time are not necessarily linked to substantial changes in profile shape, which also holds true for our extended monitoring.
Our profile variability map does pick up some low-level changes in the leading half of the brighter, secondary profile component.
However it does not appear correlated with corresponding changes in spin-down rate.

\subsubsection*{PSR J1001$-$5507 (B0959$-$54)}

Long-term monitoring of PSR~J1001$-$5507 by \citet{Chukwude2012} using the 26\,m telescope at the Hartebeesthoek Radio Astronomy Observatory (HartRAO) revealed this pulsar underwent a transition from a low to high spin-down state between MJD 48500-49300 with a $\delta\dot{\nu}/|\dot{\nu}| \sim 1.3\%$.
This coincided with the appearance of a bump in the leading edge of the pulse profile. 
The spin-down timeseries that we recover for PSR~J1001$-$5507 in Figure~\ref{fig:nudot} is initially consistent with the pulsar being in the high spin-down state, remaining somewhat flat until around MJD 58000, after which it appears to follow a linear decrease that continues to the last observation. 
On top of the secular spin-down evolution there is a clear quasi-periodic process which shows up two peaks in the Lomb-Scargle periodogram at 511\,d and 1428\,d. 
We also detect substantial changes in profile shape throughout our observing campaign, largely appearing as increased/decreased emission from the leading edge of the main peak and precursor bump.
These variations appear to be largely uncorrelated with the short-term quasi-periodic changes in spin-down rate.

\subsubsection*{PSR J1048$-$5832 (B1046$-$58)}

\begin{figure*}
    \centering
    \includegraphics[width=\linewidth]{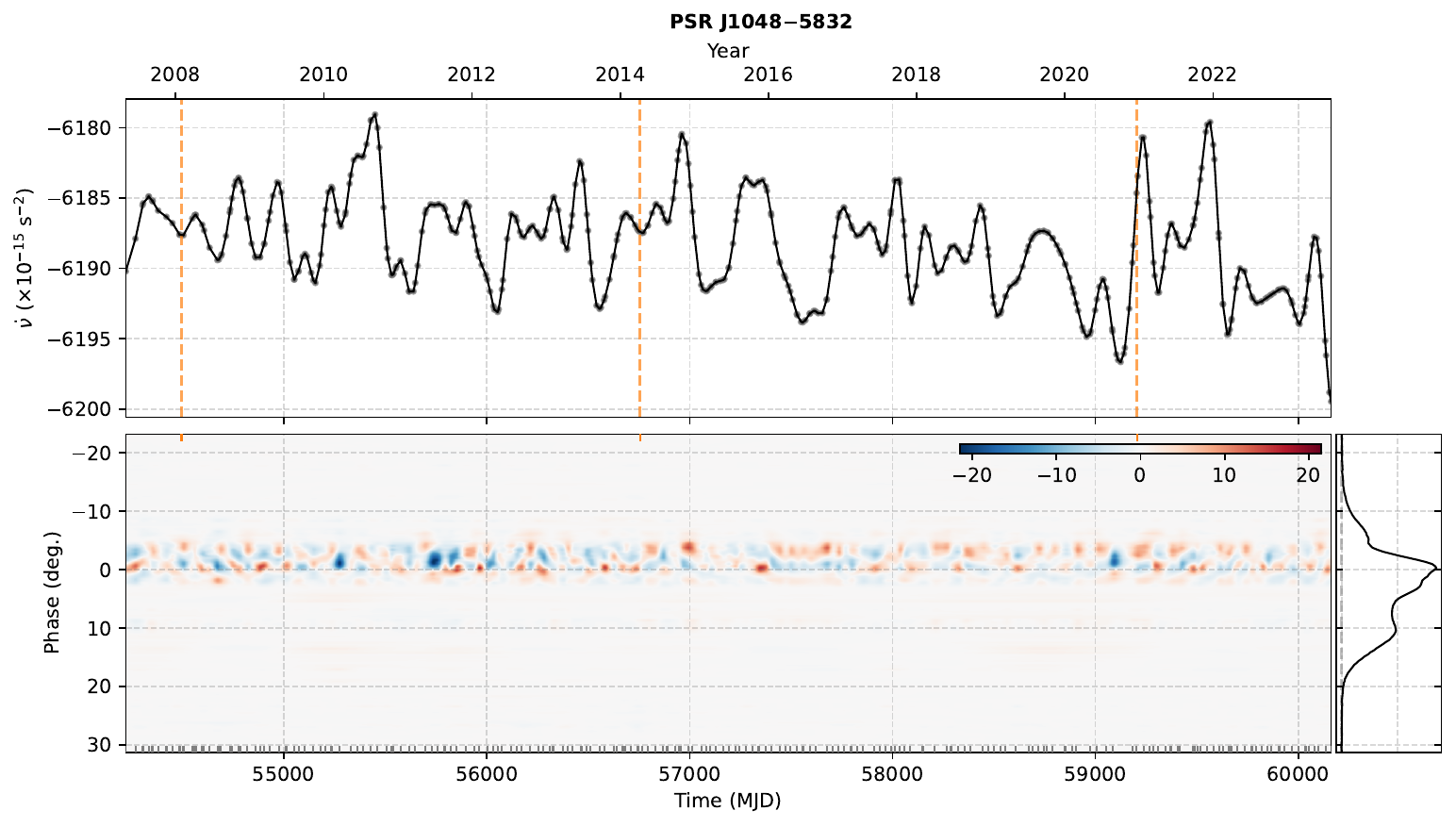}
    \caption{Rotation and emission changes in the known variable pulsar PSR~J1048$-$5832. Top panel shows the spin-down timeseries, lower panel the profile variability map with the median profile depicted on the right. Orange dashed lines in the top panel indicate the epochs of glitch events. The short orange tick on the upper part of the variability map indicates the epochs of detected glitches, while the grey ticks along the bottom are the observation epochs at which the individual profiles were obtained.}
    \label{fig:var_J1048}
\end{figure*}

This bright gamma-ray pulsar, PSR J1048$-$5832 is known to display quasi-periodic switching between a weak and strong emission state every $\sim$17 rotations \citep{Yan2020}.
Our spin-down model and profile variability map in Figure~\ref{fig:var_J1048} shows this variable emission behaviour also extends to longer timescales.
The profile variations appear confined to the leading shoulder and primary profile components, each displaying continuous evolution with time similar what is observed in the main peaks of PSRs J0908$-$4913, J1600$-$5400 and J1602$-$5100.
It is therefore not surprising that the time-averaged inter-glitch profiles highlighted by \citep{Liu2024} appear different to one another.
The spin-down variations are qualitatively complex, though a Lomb-Scargle periodogram recovers a strong quasi-period at $\sim$480\,d.
Our observed changes in the profile shape are only weakly correlated with $\dot{\nu}$ on a similar timescale.

\subsubsection*{PSR J1057$-$5226 (B1055$-$52)}

An orthogonally rotating pulsar, PSR J1057$-$5226 is known to display simultaneous, short-timescale emission state switching across both its main pulse and inter pulse \citep{Biggs1990, Weltevrede2012}.
The spin-down rate of this pulsar varies smoothly over time by $\delta\dot{\nu}/|\dot{\nu}| \sim 0.27\%$, with no clear quasi-periodicity picked up in the Lomb-Scargle periodogram.
Our profile variability map reveals that inter-observation changes in the pulse profile of this pulsar are largely restricted to the main pulse, where the brightest peak displays rapid fluctuations in intensity, while the more central component has a long-term decay in intensity superimposed on similar rapid fluctuations.
This slow variation is strongly anti-correlated with the spin-down timeseries. 

\subsubsection*{PSR J1105$-$6107}

This 63\,ms pulsar with a double-peaked profile was noted by \citet{Brook2016} as displaying a significant deficit (excess) in emission peak from the leading (trailing) peak around $\sim$MJD 56500, coinciding with an increased spin-down rate. 
Our extended dataset reveals that this event was short-lived, lasting only $\sim$200\,d.
A similar yet smaller scale profile/spin-down event occurred around MJD 57900. 
By far the largest changes in the rotation and emission characteristics of PSR~J1105$-$6107 took place between MJD 54200-55600, where a substantial deficit in emission from both profile components occurs at the same time as several large fluctuations in spin-down rate and both moderate ($\Delta\nu_{g}/\nu \sim 29.5 \times 10^{-9}$) and large ($\Delta\nu_{g}/\nu \sim 950 \times 10^{-9}$) glitches on MJDs 54711 and 55300.

\subsubsection*{PSR J1326$-$5859 (B1323$-$58)}

Timing of this pulsar at HartRAO over a 13\,yr revealed its timing residuals displayed oscillatory variations with a best-fit quasi-period of 2560\,d \citep{Frescura2003}. 
Gaussian process regression analysis of the HartRAO timing by \citet{Maritz2015} showed this behaviour can be explained by quasi-periodic spin-down variations occurring on this timescale.
Our analysis of the Murriyang observations clearly recovers the long-term quasi-periodic changes in spin-down rate of PSR~J1326$-$5859. 
However, the periodogram reveals a sharp quasi-period of only 1272\,d, approximately half the period found by \citet{Frescura2003}.
This pulsar also displays significant profile shape changes, though the high amplitude, rapid variations either side of the main peak are likely a result of profile jitter as an excess on one side of the peak is always accompanied by a deficit on the other.
The more subtle long-term changes in the precursor and tail components appear to be genuine state changes in the pulsar emission. 
The precursor component changes are weakly correlated with variations in the spin-down rate at zero-lag while changes in the tail lag the spin-down rate by $\sim 500$\,d.

Curiously, the precursor component appears to be absent in the HartRAO pulse profiles presented in \citet{Maritz2015}, which were generated by averaging data between MJD 48000-49500, yet appears in our earliest archival Murriyang observation on MJD 49589.
It is possible that a profile shape-change event in the gap occurred between reported profiles, similar to the one exhibited by PSR~J0738$-$4042 albeit without an accompanying large alteration in spin-down state \citep{Karastergiou2011, Brook2014}.
Another possibility is that the precursor is blended into the leading edge of the main profile component by dispersion measure smearing in the HartRAO data.
The high dispersion measure of 287\,pc\,cm$^{-3}$ would introduce a 3.1\,ms delay at 1.6\,GHz, which is comparable to the precursor component width.
We also note that HartRAO employed a single-polarization (right-hand circular) receiver during this period.
Differences in detected elliptical polarization at different parallactic angles may have contributed to apparent changes in profile shape.
A more detailed analysis of the HartRAO dataset may resolve this mystery.

\subsubsection*{PSR J1326$-$6700 (B1322$-$66)}

\citet{Wen2020} found that this pulsar switches between at least three distinct emission states in its single pulses. 
These include a bright mode where the profile is comprised of three overlapping sub-components, a weak state where emission from the central and trailing component are suppressed, and occasional nulls.
In terms of its long-term variability, we find PSR J1326$-$6700 displays distinct sinusoidal variations in its spin-down rate, albeit of varying amplitude, with a quasi-period of $\sim$427\,d.
These changes in spin-down are weakly correlated with deviations away from the median profile shape on the same timescale.
The shape changes are likely a result of PSR J1326$-$6700 the pulsar switching between preferentially emitting in one emission mode (bright or weak), like many of the pulsars presented by \citet{Lyne2010}.

\subsubsection*{PSR J1359$-$6038 (B1356$-$60)}

Identified by \citet{Brook2016} as a variable pulsar, the Gaussian-like profile of PSR~J1359$-$6038 often displays periods of excess emission on the right-hand slope that are correlated with transient decreases in spin-down rate that gradually decay back toward a steady state.
An extreme broadening of both the left and right hand sides of the profile coinciding with a sudden $\sim 0.14\%$ drop in spin-down occurred between MJD 56450-56650 \citep{Brook2016}.
Despite the much larger overall alteration in pulse shape, the spin-down rate change associated with this event is of comparable amplitude to those that coincide with the less substantial profile shape changes.

\subsubsection*{PSR J1401$-$6357 (B1358$-$63)}

This 0.84\,s pulsar exhibits both sub-pulse drifting and nulling, which were identified in single-pulse data collected using the Parkes UWL in 2018 \citep{Chen2023a}.
Over longer timescales, the blended two-component profile displays slow variations across the leading edge and rapid switching between excesses and deficits of emission over the peak and trailing shoulder.
Extended periods in the emission deficit state appear to coincide with two transient increases in $\dot{\nu}$ took place between MJD 57251-57887 and MJD 58891-59758, with respective peak amplitudes of $\delta\dot{\nu}/|\dot{\nu}| \sim 0.9$\% and $\sim 1.7$\%. 
This is reflected in our correlation analysis, which shows a strong positive correlation between the changes in trailing shoulder emission and the spin-down timeseries.

\subsubsection*{PSR J1600$-$5044 (B1557$-$50)}

This is another pulsar with spin-down variations that were initially identified in HartRAO observations \citep{Chukwude2003} and profile shape changes by \citet{Brook2016}.
The profile shape variations are largely confined to the leading and trailing edges, appearing quite subtle when compared to other bright variable pulsars.
As noted by \citet{Brook2016}, the largest shape change occurred at the beginning of the Parkes young pulsar monitoring project in mid-2007, appearing as a strong deficit of emission either side of the profile peak in our variability map.
The spin-down timeseries is dominated by triangular `saw-tooth' shaped variations, where an initial decrease in $\dot{\nu}$ is followed by a prolonged linear increase until the next decrease.
This is superimposed on a higher-frequency fluctuation in spin down that is not well resolved. 
The profile shape and spin-down changes are only weakly correlated.
A Lomb-Scargle periodogram formed from the spin-down timeseries shows a strong peak at 1535\,d, which matches the approximate delay between the largest spin-down events. 

\subsubsection*{PSR J1602$-$5100 (B1558$-$50)}

This is one of the best examples of a pulsar that displays a strong change in profile shape that coincides with a transient spin-down event in the Parkes data analysed by \citet{Brook2016}.
The spin-down rate of this pulsar decreased by $\delta\dot{\nu}/|\dot{\nu}| \sim 4.5\%$ over a 958\,d period between MJD 54600-55400 at the same time as a decrease in the emission from the main profile component and the appearance of two sub-peaks in the lower intensity trailing component. 
This event identified by \citet{Brook2016} remains the most intense profile shape/spin-down event to have occurred in this pulsar to date.
A similar, albeit weaker, appearance of the transient sub-peaks occurred between MJD 59005-59543 and coincided with a short decrease in pulsar spin-down rate.
Two additional spin-down events of similar amplitude to this most recent change spanning MJD 49090-49715 and MJD 51383-5200 are also readily apparent in the legacy timing data (see Figure~\ref{fig:nudot}).
The archival profiles are of insufficient quality to reliably confirm or rule out profile shape changes associated with these old events.
There is no obvious quasi-period in the spacing between subsequent events, and no such periodicity is evident in a periodogram computed from the spin-down timeseries.

\subsubsection*{PSR J1645$-$0317 (B1642$-$03)}

Cyclical variations in the timing of PSR~J1645$-$0317 were initially identified by the Jet Propulsion Laboratory pulsar timing programme \citep{Cordes1985}, which was interpreted as potential evidence the pulsar was undergoing free precession \citep{Shabanova2001}. 
Later analysis suggested this behaviour resulted from a series of `peculiar', small-amplitude ($\Delta\nu_{g}/\nu \sim 10^{-9}$) glitches that were each separated in time by $\sim$600\,d \citep{Shabanova2009}.
Like many of the bright pulsars among our sample, the variability map for PSR J1645$-$0317 displays rapid, intense variations across the main peak of its pulse profile that likely originates from pulse jitter.
However, there are also less intense, secular changes in the intensity of its precursor component.
The precursor changes are strongly correlated with quasi-periodic changes in the spin-down rate on a timescale of approximately 610\,d, with an excess of emission occurring when the pulsar is in a low spin-down state and a deficit in the high state.
A longer timescale quasi-period of $\sim$1425\,d is also apparent in the spin-down timeseries and Lomb-Scargle periodogram.
The 610\,d quasi-period almost exactly matches the average spacing between the glitch-like events reported by \citet{Shabanova2009}.
While there is a cuspy feature in our PSR~J1645$-$0317 timing residuals, the turnover is drawn out over many observations as opposed to a sharp glitch-like discontinuity.
Hence, the timing events that were previously reported as being glitches are likely to be misclassified spin-down variations.
Similar such events in other pulsars, alongside random fluctuations in pulsar spin due to timing noise, have previously been misidentified as small glitches \citep{Lower2020, Lower2023}.

\subsubsection*{PSR J1705$-$1906 (B1702$-$19)}

This is a $0.3$\,s orthogonal rotator that was found to display `on/off' sub-pulse modulation across both its the main and inter pulse \citep{Weltevrede2007}. 
Like PSR J1057$-$5226, the spin-down rate of PSR J1705$-$1906 varies slowly over time, with an apparent, dominant quasi-period of $\sim$2645\,d.
Our profile variability map does not recover any shape variations that are substantially different from fluctuations in the off-pulse noise, suggesting the rapid state switching every 10.4 rotations identified by \citet{Weltevrede2007} must be stable over long time scales.
This is unlike other state switching pulsars, where the profile shape changes recovered in the profile variability maps reflect the pulsar preferentially spending more time in one emission state than the other.

\subsubsection*{PSR J1703$-$4851}

This $1.4$\,s pulsar was noted by \citet{Johnston2021a} as switching between two distinct emission states: a `weak' state where the profile consists of three approximately equal intensity sub-components and a `bright' mode where the central component is up to an order of magnitude brighter than in the weak-state.
Our spin-down model displays small variations with time of up to $\delta\dot{\nu} \sim 0.14\%$, albeit with large uncertainties, that is only weakly correlated with changes in profile shape.
A small glitch with an amplitude of $\Delta\nu_{g}/\nu =  11.21(5) \times 10^{-9}$ and change in spin-down rate of $\Delta\dot{\nu}_{g}/\dot{\nu} = 3.0(2) \times 10^{-3}$ occurred around MJD 58570. 
Curiously, the profile variability map shows the bright state appears to have been more prevalent in observations taken before MJD 58750.
It is unclear if the two phenomena are related since the bright-mode deficit lags the glitch by $\sim 100$\,d.

\subsubsection*{PSR J1705$-$3950}

This pulsar was noted by \citet{Basu2024} as displaying alternating fluctuations in intensity between the two peaks of its profile in long-term monitoring observations with MeerKAT.
This behaviour appears to be a variant of that identified in the \citet{Lyne2010} sample, where the apparent switching between profile component intensities is due to the pulsar preferentially emitting from one component over the other (see the single-pulse data in Figure 5 of \citealt{Basu2024}).
The switching in component intensities is marginally visible in our profile variability map. 
Our spin-down timeseries reveals the pulsar initially displayed clear quasi-periodic variability with two peaks in the Lomb-Scargle periodogram at 157\,d and 314\,d.
This behaviour was seemingly interrupted by a glitch that occurred around MJD~58236, after which the quasi-periodicity becomes less defined, which may partially explain the lack of correlation with the emission state changes.

\subsubsection*{PSR J1717$-$4054}

Situated somewhere between nulling and intermittent pulsars, PSR~J1717$-$4054 exhibits three distinct null states lasting a few to tens of rotations during `active states', and long inactive states that last many thousands of rotations \citep{Kerr2014}.
As a result of its quasi-intermittency, we have comparatively few detections of this pulsar throughout our timing programme which results in the Gaussian process regression fits to its timing and profile residuals returning largely stochastic variations in both.
There is a weak anti-correlation between the profile variability and the spin-down model.
The $\delta\dot{\nu}/|\dot{\nu}| \sim 7.91\%$ displayed by this pulsar is the largest among our sample.

\begin{figure*}
    \centering
    \includegraphics[width=\linewidth]{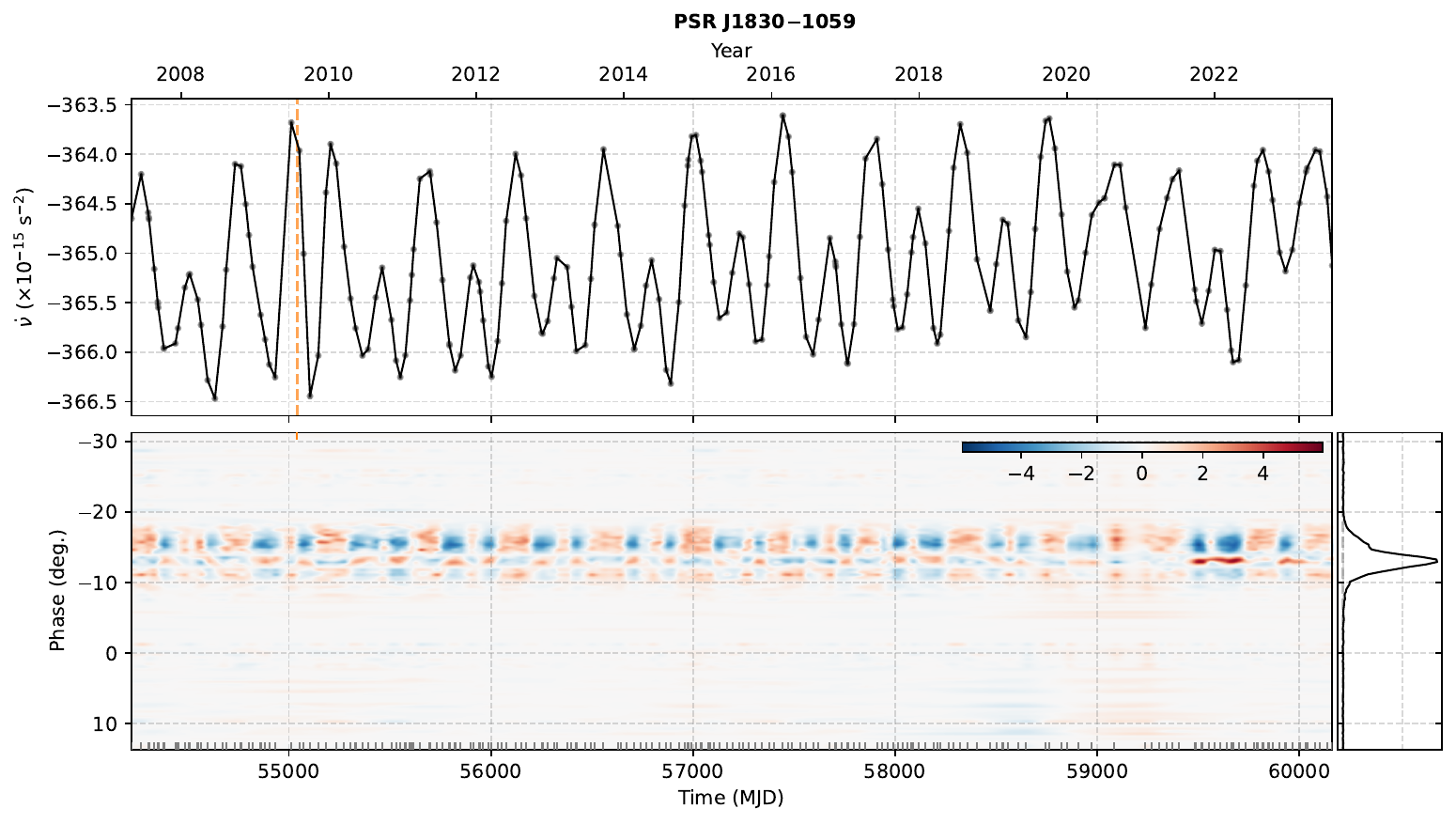}
    \caption{Spin-down timeseries and profile variability map for PSR~J1830$-$1059. As Figure~\ref{fig:var_J1048} otherwise.}
    \label{fig:var_J1830}
\end{figure*}

\subsubsection*{PSR J1741$-$3927 (B1737$-$39)}

The main peak and trailing shoulder components of this pulsar were noted by \citet{Basu2024} as displaying correlated fluctuations in intensity, where an increase in the emission peak corresponded to a deficit in trailing shoulder emission, and vice-versa.
No spin-down variability was reported in their analysis of PSR~J1741$-$3927, however this appears to have been a result of the short timing baseline of the MeerKAT Thousand Pulsar Array, as our $\dot{\nu}$ timeseries shows a clear long-term fluctuation over our 10\,yr of Murriyang observations.
The profile variability map that we generated also shows the main peak undergoing clear, rapid variations on a timescale of a few hundred days, though we do not see similar anti-correlated changes in the trailing shoulder component.
Instead, both the leading and trailing shoulders appear to undergo a secular increase in intensity that is correlated with the long-term change in $\dot{\nu}$ at a 1000\,d lag.

\subsubsection*{PSR J1825$-$0935 (B1822$-$09)}

A 0.77\,s pulsar, PSR~J1825$-$0935 is a well studied emission/spin-down mode-switching pulsar. 
Its radio emission switches between a `bright mode' where a precursor component in the main pulse becomes active and the inter-pulse is suppressed, and a `quiet' mode where the precursor component is inactive and the inter-pulse is active \citep{Fowler1981}.
Single-pulse studies have shown that the switching in the main- and inter-pulse components occurs simultaneously indicating a strong degree of information exchange between the emission above both magnetic poles \citep{Fowler1982}.
Unlike several other mode-changing pulsars, the emission state-switching does not appear to extend to X-ray energies \citep{Hermsen2017}.
\citet{Lyne2010} demonstrated that previously reported `slow glitches' in the timing of PSR~J1825$-$0935 \citep{Shabanova2005} corresponded to brief periods where the pulsar was in a low-$\dot{\nu}$ state and preferentially emitted in the bright mode.
Our Parkes timing of PSR~J1825$-$0935 only covers the last spin-down event described in \citet{Lyne2010}, where the spin-down decreased by $\delta\dot{\nu}/|\dot{\nu}| \sim 2.7\%$.
This event does not overlap with our profile variability map.
There is also a gap in our timing between MJD 54144-54762, immediately following a moderately sized ($\Delta\nu_{g}/\nu \sim 1.17 \times 10^{-7}$; \citealt{Lower2021b}) glitch on MJD 54115.
Three small events with respective fractional decreases in spin-down of $\delta\dot{\nu}/|\dot{\nu}| \sim 0.6\%$, $\sim 0.4\%$ and $\sim 1.1\%$ respectively that took place over MJD 57881-58222, MJD 58358-58669 and MJD 59344-59640.
The last of these events was followed by an enhancement in the spin-down rate by $\delta\dot{\nu}/|\dot{\nu}| \sim 0.3\%$ between MJD 59640-60131 that coincided with the pulsar spending an extended period of time in the quiet emission mode, a first for PSR~J1825$-$0935.

\subsubsection*{PSR J1830$-$1059 (B1828$-$11)}

This is one of the best known mode-changing pulsars that displays highly correlated emission/profile variations. 
Initially thought to be evidence of free precession \citep{Stairs2000}, PSR~J1830$-$1059 smoothly varies between three distinct spin-down states (high, medium and low, see Figure~\ref{fig:var_J1830}) with a min/max $\delta\dot{\nu}/|\dot{\nu}| \sim 0.8\%$ on timescales of approximately 500 and 220\,d.
\citet{Lyne2010} showed these changes in spin-down were correlated with changes in the profile shape, later revealed to be due to the pulsar spending preferentially more time in one of two extreme pulse shapes: a single bright component or a two-component profile consisting of a precursor bump at the leading edge of a bright peak \citep{Stairs2019}.
Both the highly quasi-periodic spin-down rate and profile shape variations are easily recovered in our Gaussian process analysis of the Parkes data, which we show in Figure~\ref{fig:var_J1830}.
A small $\Delta\nu_{g}/\nu \sim 6.3 \times 10^{-9}$ glitch was reported to have occurred on MJD 55040 \citep{Basu2022}, which lines up with a period in our spin-down timeseries where a large decrease in $\dot{\nu}$ occurred instead of an anticipated small change.
Curiously, the modulation pattern in the spin-down timeseries appears to have changed after $\sim$MJD 59100.
Instead of alternating between large and small changes in spin-down every 221\,d, the pattern has become somewhat chaotic. 
Two large decreases in spin down occurred without an intervening small change, followed by a small change, and then two large changes. 
At the same time, the radio profile spent extended periods of time in the two or single component states.  
While the more recent profile variations appear to have recovered back to the pre-MJD 59100 state, it is unclear yet whether the spin-down modulation has returned back to its `normal' alternating between large/small spin-down state changes.

\subsection{New pulsars with both profile and spin-down variations}

Alongside the known variable pulsars, we have discovered another 28 pulsars that show fluctuations in both profile shape and spin-down rate.
Here we summarise the kind of variations detected in the radio emission and spin-down properties of these pulsars.

\subsubsection*{PSR J0134$-$2937}

This pulsar displays a slow, smooth change in $\dot{\nu}$ by $\sim$0.01\% over our 16\,yr of Parkes monitoring.
It also displays significant profile variability that is confined to the peak and trailing edge.
These shape changes are uncorrelated with the spin-down timeseries, occurring on a substantially shorter timescale than the corresponding changes in $\dot{\nu}$.
We note that this pulsar does exhibit a substantial amount of diffractive scintillation due to multi-path propagation in the interstellar medium.
However, apparent scintillation-induced shape variations would affect the entire profile, as opposed to the relatively resitricted region of the profile that we observe to vary.

\subsubsection*{PSR J0255$-$5304 (B0254$-$53)}

The radio profile of this pulsar is has two peaks that we found to exhibit substantial variability.
The leading component shows both fast and slow changes, while the trailing component only varies slowly with time.
In contrast, the spin-down rate varies smoothly over time, with a potential quasi-period of $\sim$1648\,d.
Our correlation map suggests that the profile variations lag the changing spin-down rate by 500-1000\,d.

\subsubsection*{PSR J0543$+$2329 (B0540$+$23)}

This pulsar displays rapid changes in both profile shape and spin-down rate.
Our profile variability map indicated the shape variations are largely due to longitudinal shifts in the central part of the profile that are not strongly correlated with changing spin-down.
The periodogram computed from the spin-down timeseries has two peaks either side of 1\,yr, which may be an artifact of a position offset in the timing model.

\subsubsection*{PSR J0630$-$2834 (B0628$-$28)}

This pulsar displays a broadly Gaussian shaped radio profile, albeit with many blended sub-components.
Its spin-down timeseries in Figure~\ref{fig:nudot} exhibits slow $\delta\dot{\nu}/|\dot{\nu}| \sim 0.72$\% fluctuations over 23.5\,yr of timing. 
This is contrasted by the profile variability map, which shows rapid changes across much of the profile peak that appear somewhat uniform across the smaller sub-components. 
Our cross-correlation analysis returns no substantial positive or negative correlation between the profile variability map and the spin-down timeseries.

\subsubsection*{PSR J0631$+$1036}

The profile of this pulsar is comprised of four sub-components: two dominant peaks that are flanked by weaker leading and trailing components.
Our derived profile variability map reveals small, transient changes in the two dominant components that appear as a deficit in the leading bright peak and an excess in the trailing peak.
These shape changes appear to coincide with short-duration decreases in spin-down rate.
However, there is no strong correlation between the variability map and $\dot{\nu}$ timeseries, which displays seemingly chaotic, short-term variations on top of a longer-term linear decrease. 
The Lomb-Scargle periodogram generated from the $\dot{\nu}$ model fails to recover a substantial quasi-period. 

\subsubsection*{PSR J0729$-$1836 (B0727$-$18)}

This pulsar switches between two quasi-stable spin-down states with $\delta\dot{\nu}/|\dot{\nu}| \sim 0.54\%$, similar to what is seen in PSR~J2043$+$2740 though without clear correlated changes in profile shape \citep{Lyne2010, Shaw2022}.
Its profile shape variations are confined to the leading peak and trailing edge of the profile.
However, unlike the variations seen in PSR~J2043$+$2740, our correlation map shows these changes in emission are only weakly anti-correlated with the switches between spin-down states.

\subsubsection*{PSR J0820$-$1350 (B0818$-$13)}

This is a slower 1.3\,s pulsar, PSR J0820$-$1350 displays a blended double-peaked profile and a bump in the leading edge.
The secondary peak has a substantial amount of rapid variability with time. 
These rapid profile variations are in contrast with the slow shifts in spin-down rate by $\delta\dot{\nu}/|\dot{\nu}| \sim 0.21$\% over time. 
Our Lomb-Scargle periodogram analysis recovers a possible quasi-period of $\sim$1281\,d in the spin-down timeseries for this pulsar. 

\subsubsection*{PSR J0942$-$5552 (B0940$-$55)}

This pulsar displays correlated profile and spin-down variations that fluctuate on relatively long timescales of 2-3 years.
We find profile shape changes occurring in the leading precursor component, right-hand shoulder of the main peak and the trailing component. 
Excess emission from these regions are observed when the pulsar is in a lower spin-down state.
Small deficits of emission occur when the spin-down rate is higher.
Our periodicity analysis detects a substantial amount of power at a 1\,yr quasi-period, which likely results from a small positional offset in the timing model from the true sky location.

\subsubsection*{PSR J1056$-$6258 (B1054$-$62)}

Similar to PSR J0742$-$2822, this pulsar displays a complicated relationship between its highly variable profile shape and spin-down rate.
The spin-down rate shows rapid changes with time, with quasi-periods of 456\,d and 177\,d.
The entire profile is affected by shape variations that appear strongest across the peak, which also vary on short timescales.
Emission and spin-down processes appear to be uncorrelated with one another similar to the pre-glitch behaviour of PSR~J0742$-$2822.

\subsubsection*{PSR J1224$-$6407 (B1221$-$63)}

This is a bright, 0.22\,s radio pulsar that displays two strong quasi-periods in its long-term spin down rate at $\sim$427\,d and $\sim$747\,d.
Our profile variability map shows rapid changes across much of the profile on timescales that are shorter than both detected quasi-periods in the spin-down timeseries, indicating the two processes are largely decoupled from one another.
The strongest of these variations occur in the precursor component, which appear uncorrelated with the spin-down rate, as opposed to variations in the profile peak which are mildly correlated with $\dot{\nu}$.

\subsubsection*{PSR J1243$-$6423 (B1240$-$64)}

\begin{figure}
    \centering
    \includegraphics[width=\linewidth]{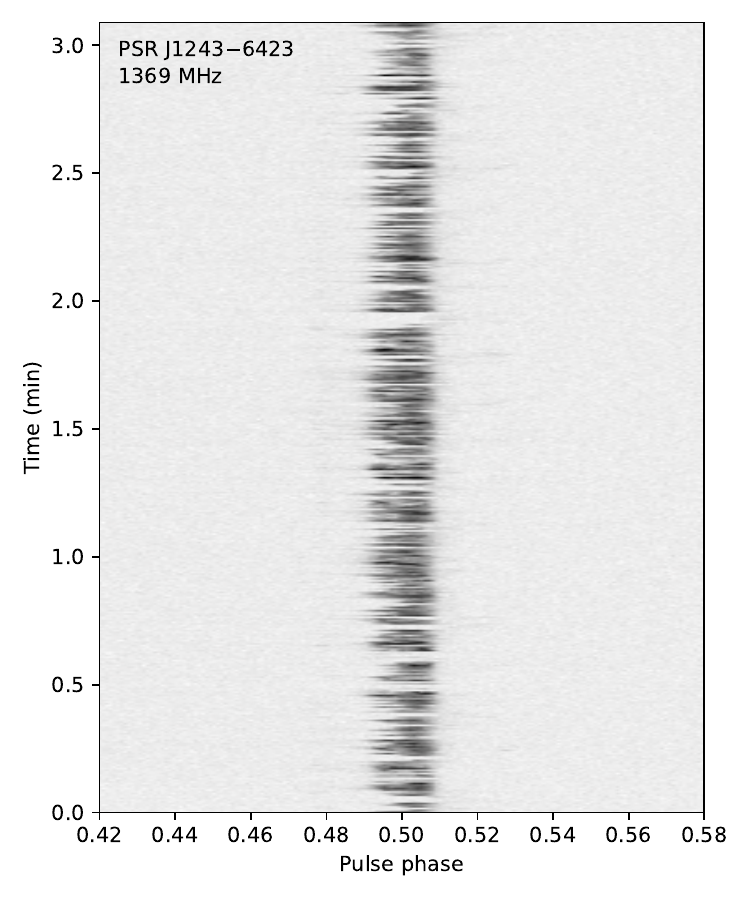}
    \caption{Stack of 482 single pulses from PSR~J1243$-$6423 recorded using the Parkes 20-cm multibeam receiver on MJD 57600. Nulls and weak radio pulses are clearly visible as gaps among the normally bright emission from this pulsar.}
    \label{fig:j1243_sp}
\end{figure}

Similar to PSR~J1224$-$6407, this pulsar displays substantial profile shape variations on a more rapid timescale than the observed changes in spin-down rate.
Our correlation analysis indicates there is a $\sim$200\,d lag between changes in $\dot{\nu}$ and the bump in the main profile component changes, and a $\sim$400\,d lag with changes in the trailing edge.
The spin-down timeseries displays two distinct quasi periods at $\sim$626\,d and $\sim$1203\,d.
Figure~\ref{fig:j1243_sp} shows a set of archival single-pulse data collected by Murriyang on MJD~57600, which reveals the radio emission of PSR~J1243$-$6423 switches between at least two distinct states. 
This includes a bright state, where emission is detected from across the profile, and a weak state where the emission primarily originates from a low-flux density trailing profile component. 
There are also nulls in between several switches between emission state.
This nulling behaviour was previously identified in single pulse data collected by the Molonglo Observatory Synthesis Telescope \citep{Biggs1986}.
Comparing the on-pulse flux density against the root-mean-square (RMS) of the off-pulse region in Figure~\ref{fig:j1243_sp} returned an approximate nulling fraction of 0.12, a factor of three higher than the previously reported value of $\leq 0.04$ \citep{Biggs1992}.
This could be due to a preference for null pulses when PSR J1243$-$6423 is in a particular spin-down state, similar to what is seen among other emission state-switching pulsars.

\subsubsection*{PSR J1327$-$6222 (B1323$-$62)}

PSR J1327$-$6222 is another pulsar that displays rapid, albeit seemingly uncorrelated, variations in both its radio profile shape and spin-down rate over time.
The profile shape changes appear in both components of the blended, two-peaked profile with no clear quasi-stable states that the pulsar switches between.
Given the rapid changes in both profile shape and spin-down, it is possible our approximately monthly observing cadence is unable to fully resolve any underlying quasi-stable states.
These somewhat random profile shape changes are interspersed with brief periods of enhanced emission from a bump in the trailing shoulder, the two most prominent of which occurred around MJD 58915 and MJD 59768.

\subsubsection*{PSR J1328$-$4357 (B1325$-$43)}

This is a pulsar with a blended two-component profile with shape variations that are predominantly restricted to increased/decreased emission from the trailing component.
These shape variations are uncorrelated with the changing spin-down rate, which has a dominant quasi-period that peaks near 365\,d.
Given the somewhat ragged appearing spin-down timeseries, this quasi-period may be indicative of a strong degeneracy between an incorrectly modelled pulsar position with a near-annual quasi-periodic process.

\subsubsection*{PSR J1428$-$5530 (B1424$-$55)}

This pulsar displays slow changes in spin-down rate with an amplitude of $\delta\dot{\nu}/|\dot{\nu}| \sim 0.12$\%.
However its variable profile shape fluctuates on a much shorter month-long timescales.
The observed profile variations appear as increases in width and are most prominent in the trailing edge of the overall Gaussian shape. 
These changes in the trailing edge are moderately correlated with the spin-down timeseries.
Inspection of the profile variability map shows the enhanced emission from this part of the profile appears more prevalent with the pulsar is in a lower spin-down state.

\subsubsection*{PSR J1453$-$6413 (B1449$-$64)}

This is a 0.18\,s pulsar that has undergone four glitches since the beginning of our timing programme in 2007. 
The spin-down timeseries of PSR J1453$-$6413 is largely flat, except for two noteworthy fluctuations in $\dot{\nu}$, one that took place between MJD 57057-58645 and another that began around MJD 59182 and is ongoing.
These events appear to be associated with excess emission from the leading bump in the pulse profile, though our correlation analysis suggests there is only a weak positive correlation between the variability map at these pulse phases and the spin-down timeseries.
The more rapid profile shape changes that occur in the leading edge of the profile peak are uncorrelated with the changes in $\dot{\nu}$.
It is unclear if these profile/spin-down events are related to the small amplitude glitches that took place several hundred days prior to their onset.

\subsubsection*{PSR J1456$-$6843 (B1451$-$68)}

PSR J1456$-$6843 is an example of a pulsar where changes in profile shape are not associated with variations in $\dot{\nu}$.
It has a complicated, albeit symmetric pulse profile that appears comprised of multiple blended components.
Our variability map shows these components each varying randomly over time, which according to the correlation map is decoupled from the substantially slower $\delta\dot{\nu}/|\dot{\nu}| \sim 0.75$\% change in spin-down rate over time.

\subsubsection*{PSR J1559$-$4438 (B1556$-$44)}

This pulsar has a triple-component radio profile, with low intensity precursor and trailing components and a bright central component that has a `shoulder` on its trailing edge.
All three components display shape variations different timescales.
The precursor and trailing components both show simultaneous excess and suppressed emission that is correlated with the changing spin-down rate of the pulsar.
Emission from the leading edge of the central component varies on a similar timescale, but lags changes in the precursor/trailing components by $\sim$400\,d.
The shoulder on the central component displays more rapid variations, yet appears to preferentially emit in a weak state in concert with the precursor and trailing components.
Changes across the central component peak are anti-correlated with the varying spin-down rate and are out of phase with changes in the other components by half a cycle.
While the spin-down rate of PSR J1559$-$4438 visually appears to display both long and short timescale variations, with the more rapid changes appearing quasi-periodic, no substantial peaks are detected in the corresponding Lomb-Scargle periodogram.

\subsubsection*{PSR J1604$-$4909 (B1600$-$49)}

Another triple-component pulsar, where the changes in intensity displayed by the bright central peak are highly correlated with step-like changes in spin-down rate.
The shape changes are characterised by the appearance of an excess bump in the central component peak whenever the pulsar is in a low spin-down rate, which rapidly disappears over the $\sim$200\,d transition to the high spin-down state.
This is somewhat reminiscent of the transient variations reported in PSR~J2043$+$2740 \citep{Lyne2010, Shaw2022}, however the changes in PSR~J1604$-$4909 display a $\sim$1698\,d cycle between emission/spin-down states.
While there is an apparent shorter timescale process that is superimposed on the spin-down timeseries, our Lomb-Scargle periodogram does not recover any significant power at periods other than that associated with the long-term process.

\subsubsection*{PSR J1703$-$3241 (B1700$-$32)}

The spin-down rate for this pulsar displays a gradual decrease over eight years of Parkes monitoring.
This is in contrast to the profile variability map, which shows rapid changes in emission across its two dominant sub-components, with the brighter leading peak showing the strongest intensity fluctuations.
As expected, these changes in profile shape are uncorrelated with the decrease in $\dot{\nu}$.

\subsubsection*{PSR J1709$-$1640 (B1706$-$16)}

This 0.65\,s pulsar has a relatively simple radio profile that displays substantial fluctuations in intensity across its trailing edge.
These changes in profile shape appear to be weakly correlated with variations in $\dot{\nu}$, albeit with a $\sim 100$\,d lag.
The Lomb-Scargle periodogram computed from the spin-down timeseries displays a strong peak at 359\,d, which could be a result of a near-1\,yr quasi-period beating with a small discrepancy in the pulsar position in the timing model.

\subsubsection*{PSR J1745$-$3040 (B1742$-$30)}

\begin{figure}
    \centering
    \includegraphics[width=\linewidth]{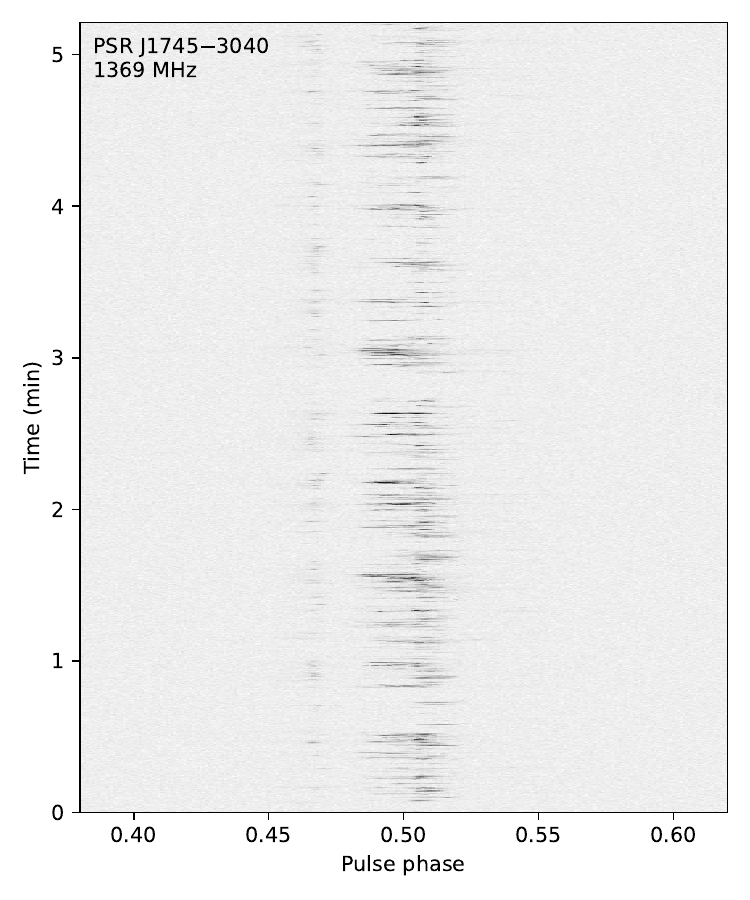}
    \caption{A set of 850 single pulses from PSR~J1745$-$3040 recorded using the Parkes 20-cm multibeam receiver on MJD 56466. The `off' emission state appears as gaps in the detected radio emission.}
    \label{fig:j1745_sp}
\end{figure}

PSR~J1745$-$3040 is a bright gamma-ray pulsar with a 0.37\,s spin period.
It has a two-component radio profile with the brighter trailing component consisting of a bright peak and a leading bump.
Both components display substantial epoch-to-epoch changes in intensity in the profile variability map that are uncorrelated with the slow, secular changes in $\dot{\nu}$ over time.  
This variability appears to be a result of both our short observation times and emission state switching in the single pulses of PSR~J1745$-$3040, behaviour that is also seen in similarly bright pulsars such as PSRs J1048$-$5832 and J1243$-$6423.
In Figure~\ref{fig:j1745_sp} we present a stack of single pulses detected by Murriyang, where the pulsar clearly switches between on and off emission states.
When the pulsar is in the `on` state, we detect radio emission from both profile components, occasionally interspersed with very bright bursts from the main peak of the second component.
Whereas the `off` state appears to be extended nulls, where no radio emission is detected down to the sensitivity limit of the instrument.
A simple comparison between the mean on-pulse flux density and the off-pulse RMS reveals an approximate nulling fraction of 0.51 during the observation presented in Figure~\ref{fig:j1745_sp}.
Unlike the other pulsars among our sample that display similar behaviour, there does not appear to be a strong association between for the pulsar preferentially emitting in one emission state over the other and the changes in $\dot{\nu}$ over time.
This may however, simply be due to the much lower frequency fluctuation timescale of $\dot{\nu}$ and relatively high nulling fraction of this pulsar.

\subsubsection*{PSR J1751$-$4657 (B1747$-$46)}

The radio profile of this pulsar consists of a blended, two-component profile that displays rapid fluctuations in intensity confined to the brighter, leading peak.
These intensity changes are mildly correlated with the slowly varying spin-down rate. 
There is a persistent emission deficit that coincided with the pulsar being in a high spin-down state between MJD 58122-59248.

\subsubsection*{PSR J1752$-$2806 (B1749$-$28)}

Similar to PSR J0742$-$2822, this pulsar also displays both rapid variations in profile shape and spin-down rate with a low Q-factor quasi-period of $\sim$157\,d.
The shape changes affect the entire profile to some degree, but are strongest at the peak.
Some of this variability may therefore be the result of profile jitter.
However, the shape changes are correlated with the rapidly varying spin-down rate, albeit with a $\sim$50\,d lag.
No glitches have been detected in the timing of PSR J1752$-$2806 to date.

\subsubsection*{PSR J1820$-$0427 (B1818$-$04)}

The profile of PSR J1820$-$0427 is relatively simple, consisting of a bright peak with both leading and trailing shoulders of emission.
Our profile variability map reveals changes in intensity within the leading shoulder and peak that are uncorrelated with the detected fluctuations in the pulsar spin-down rate.
This is despite the transient increase in spin-down rate of $\delta\dot{\nu}/\nu \sim 0.7$\% that occurred between MJD 58169-59088. 

\subsubsection*{PSR J1824$-$1945 (B1821$-$19)}

The radio profile of PSR~J1824$-$1945 displays rapid shape variations in spite of a slowly changing spin-down rate.
Much of the rapid changes either side of the profile peak can be attributed to epoch-to-epoch jitter, with a deficit of emission on one side being countered by an excess on the other.
There is however a clear correlation between the slowed intensity changes across the profile peak itself and the spin-down rate at a lag of $\sim$1000\,d.
Inspection of the $\dot{\nu}$ model in Figure~\ref{fig:nudot} reveals two slow, transient increases in spin-down rate that took place between MJD 53616-54441 and MJD 57564-58852 with $\delta\dot{\nu}/|\dot{\nu}| \sim 0.3$\% and $\sim 0.2$\% respectively.
A small offset in the pulsar position in our timing model is likely to be responsible for the high frequency fluctuation in $\dot{\nu}$ that is superimposed on the slow variations with time.

\subsubsection*{PSR J1829$-$1751 (B1826$-$17)}

This pulsar displays a triple peaked profile where all three components have approximately equal intensity.
Our profile variability map shows that these components all display varying levels of intensity fluctuations, with the leading and trailing components mirroring one another.
The spin-down timeseries reveals a quasi-periodic switching between high and low spin-down states every $\sim$807\,d, which is weakly anti-correlated with changes in the trailing profile component.
Visually both the leading and trailing components display an excess of emission when PSR J1829$-$1751 is in the high spin-down state, and a deficit when in the low state.
A higher-fluctuation frequency quasi-period in the spin-down rate may be linked to the peak in periodogram power at 1\,yr, likely resulting from an incorrect position in the timing model.

\subsubsection*{PSR J1845$-$0434 (B1842$-$04)}

The 0.49\,s radio pulsar PSR J1845$-$0434 displays a two-component profile consisting of a bright leading peak and a trailing shoulder.
Its spin-down timeseries reveals it undergoes transient decreases in spin-down rate by $\delta\dot{\nu}/|\dot{\nu}| \sim 0.1$-$0.9$\% that last 228-975\,d. 
Each event is separated in time by approximately 1000-2000\,d.
The first and fourth decreases in spin-down rate appear to coincide with an excess of emission from the trailing shoulder component, despite the overall profile variability map covering this segment being only weakly correlated with the $\dot{\nu}$ timeseries. 
Curiously, the recovery stage of the second spin-down event that we detect appears to also coincide with an excess in emission from the shoulder component, while the third and fifth events do not have any corresponding profile shape change.

\subsubsection*{PSR J1913$-$0440 (B1911$-$04)}

PSR~J1913$-$0440 is another bright radio pulsar with profile shape changes confined to the trailing edge of its profile.
Its changes appear to be weakly anti-correlated with high fluctuation frequency variations in the spin-down timeseries, for which our periodicity analysis failed to recover a corresponding (quasi-)period. 
Instead only the lower frequency variation with a quasi-period of $\sim$1066\,d was recovered.
Similar to PSR~J1327$-$6222, our monthly observing cadence may be insufficient to fully resolve the short-time scale quasi-periodicity.

\subsubsection*{PSR J2048$-$1616 (B2045$-$16)}

This bright 22\,mJy pulsar (at 1.4 GHz) with a triple component profile, where all three components display seemingly random fluctuations in relative intensity.
Like many of our newly identified profile/spin-down varying pulsars, the component shape changes take place on a more rapid timescale compared to the slowly changing value of $\dot{\nu}$.
However, there is a weak anti-correlation between the variability map and the spin-down rate across all three components.
The Lomb-Scargle periodogram computed from the $\dot{\nu}$ timeseries recovered a quasi-period of $\sim$1015\,d.

\subsection{Other noteworthy pulsars}

\subsubsection*{PSR J1513$-$5908 (B1509$-$58)}

An unusual young, energetic pulsar surrounded by a wind nebula in the supernova remnant SNR G320.4$-$1.2 (see, \citealt{Gaensler2002} and references therein).
In spite of its relative youth, the long-term timing of PSR J1513$-$5908 is uncharacteristically stable, enabling measurements of both its braking index of $n = 2.832 \pm 0.003$ and higher-order braking index of $m = 17.6 \pm 1.9$ \citep{Livingstone2011}.
It has also not glitched in over 30 years of timing \citep{Parthasarathy2020}, counter to the expected glitch rate of 0.9\,yr$^{-1}$ for pulsars with similar rotational properties \citep{Lower2021b}.
After correcting for the spin-frequency of PSR J1513$-$5908 and its first, second and third derivatives, our Gaussian process fit to its timing residuals reveals short timescale fluctuations in $\dot{\nu}$ with a low Q-factor quasi-period of $\sim$307\,d.
This short-term behaviour is superimposed on a longer timescale, smooth change in spin-down, where the amplitude of $\delta\dot{\nu}/|\dot{\nu}| \sim 0.03$\% is the smallest out of our entire sample.
No changes in profile shape were identified, owing to our relatively short integration times and its weak profile.

\subsubsection*{PSR J1734$-$3333}

This high magnetic field strength pulsar with an unusually low braking index of $n = 1.2 \pm 0.2$ \citep{Espinoza2011b, Lower2021b}.
Subtracting the linear slope induced by the $\ddot{\nu}$ term leaves a predominantly flat spin-down timeseries for most of its 1997-2014 timing history.
However, there are several transient changes in $\dot{\nu}$ occurring between MJD 53489-53729, MJD 54270-54550 and MJD 55425-55597.
The second and largest event appears qualitatively similar in direction and shape to a recent transient spin-down change identified in X-ray timing of PSR~J0540$-$6919 (B0540$-$69; \citealt{Espinoza2024}), albeit with an order of magnitude larger fractional increase in spin-down rate of $\delta\dot{\nu}/|\dot{\nu}| \sim 0.1\%$.
No profile shape changes are detected alongside any of these spin-down events. 
However, this is not surprising given the relatively weak and scatter-broadened profile of PSR~J1734$-$3333 at 1.4\,GHz.

\subsubsection*{PSR J1806$-$2125}

This pulsar underwent a giant glitch sometime between MJD 50850-52250 \citep{Hobbs2002}.
Although this event occurred prior to our first Murriyang timing measurements, we clearly detected the exponential recovery from it in the pulsar timing residuals.
Fitting for only the decaying component of the glitch using {\sc tempo2} at a nominal glitch epoch of MJD 51708, we inferred a $\Delta\nu_{d}/\nu = 402(27) \times 10^{-9}$ with an exponential recovery timescale of approximately 490\,d. 
While no further glitch events have occurred in this pulsar over the preceding 22\,yr of timing, its spin-down rate displays several transient decreases by about $\delta\dot{\nu}/|\dot{\nu}| \sim 0.1$-$0.3\%$ between MJDs 54379-55051, 55372-55703, 56220-56837 and MJD 59737 onwards.
A much larger transient spin-down event took place over MJD 57536-59079, where the pulsar was spinning down $0.9\%$ slower than average. 
Curiously, all four of the events detected in full share a similar morphology, consisting of a rapid initial decrease in spin-down rate, followed by a near linear increase back to the `steady' value.
A higher-frequency quasi-period appears super-imposed on top of these events.
Both the low $\sim 0.8$\,mJy flux density of PSR~1806$-$2125 at 1.4\,GHz and short per-epoch observations preclude us from detecting clear changes in profile shape associated with variations in spin-down.

\section{Discussion}\label{sec:disc}

We have identified 238 pulsars that exhibit substantial variations in their spin-down rates, of which 52 also display changes in profile shape, meaning that we have assembled the largest sample of variable pulsars to date.
The propensity of these behaviours confirms previous suggestions that radio emission and spin-down state switching is commonplace among the broader pulsar population.
With this large sample on hand we can also investigate how the amplitude of these effects scales with the average pulsar spin and spin-down rate, in addition to exploring the origins of specific sub-types of profile/spin-down rate variability.

\subsection{Scaling relations across the population}

\begin{figure}
    \centering
    \includegraphics[width=\linewidth]{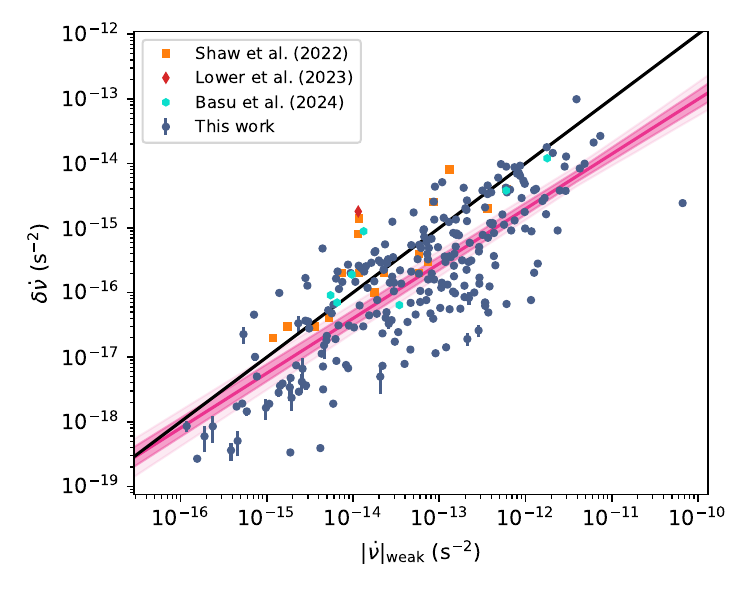}
    \caption{Relationship between spin-down variability and weak spin-down state. Values for the pulsars studied by \citet{Shaw2022} are shown in orange, the red diamond indicates PSR~J0738$-$4042 \citep{Lower2023}, pulsars from \citet{Basu2024} are the teal hexagons, and our measurements are plotted in blue. Note, there are several pulsars in common between these datasets as indicated by overlapping points. The black line is the $\delta\dot{\nu} \propto 0.01 |\dot{\nu}|_{\rm weak}$ relation from \citet{Lyne2010}. The pink line is the median a-posteriori fit to the blue points with the pink shading indicating the 68\% and 97\% confidence intervals.}
    \label{fig:nudot_pop}
\end{figure}

While a direct causal relationship between the observed spin-down variability in pulsars and external or internal physical processes is yet to be established, we can explore how this behaviour scales phenomenologically with various pulsar properties.
In Figure~\ref{fig:nudot_pop} we present the measured min/max difference in spin-down rate against the corresponding `weak' (minimum) spin-down rate value for the 238 variable pulsars in our sample, alongside values from several previous works.
\citet{Lyne2010} noted that their sample of 17 correlated profile/spin-down state switching pulsars appeared to follow a simple power-law relation, where $\delta\dot{\nu} \propto 0.01 |\dot{\nu}_{\rm weak}|$. 
A somewhat shallower power-law relation of $\log_{10}(\delta\dot{\nu}/|\dot{\nu}_{\rm weak}|) \propto 0.84 \pm 0.02 $ was recovered using extended timing of the same pulsar sample by \citet{Shaw2022}.
While there is a substantial number of pulsars in Figure~\ref{fig:nudot_pop} that reside about the original \citet{Lyne2010} relation (shown in black), about half of our larger sample are situated between one to three orders of magnitude below the relation.
This is particularly true for pulsars with both higher and lower weak spin-down rates than those among the \citet{Lyne2010} sample.

Drawing on previous works to study similar relations between pulsar properties and stochastic timing noise strength \citep{Dewey1989, Hobbs2010, Shannon2010, Parthasarathy2019, Lower2020}, we compared how $\delta\dot{\nu}$ scales with $|\dot{\nu}_{\rm weak}|$ through fitting a power-law of the form
\begin{equation}\label{eqn:nudot_scale}
    \delta\dot{\nu} = 10^{\xi}\, |\dot{\nu}_{\rm weak}|^{b},
\end{equation}
where $\xi$ is an arbitrary scaling parameter and $b$ is the scaling index of $\dot{\nu}$.
We determined the posterior distributions for the model parameters using Bayesian parameter estimation with a Gaussian likelihood function of the form
\begin{equation}
    \mathcal{L}(d | \theta) = \prod_{i}^{N_{\rm psr}} \frac{1}{\sqrt{2\pi\sigma_{i}^{2}}} \exp \Bigg[-\frac{(d_{i} - \mu(\theta))^{2}}{2\sigma_{i}^{2}} \Bigg],
\end{equation}
where $d$ is the data, $\mu$ is the power-law model, $\theta$ is the model parameters, $\sigma_{i}^{2} = \sigma_{\delta\dot{\nu},i}^{2} + \sigma_{Q}^{2}$ is the per-pulsar uncertainties on $\dot{\nu}$ with an additional error in quadrature term added to account for scatter in our measurements.
The model parameters were sampled using {\sc Bilby} as a wrapper for the {\sc dynesty} nested sampling algorithm \citep{Ashton2019, Speagle2020}.
The resulting relationship is represented by the pink line in Figure~\ref{fig:nudot_pop}, where
\begin{equation}\label{eqn:f1_relation}
    \delta\dot{\nu} = 10^{-4.5 \pm 0.5}\,|\dot{\nu}|_{\rm weak}^{0.85 \pm 0.04}.
\end{equation}
This is shallower than the aforementioned relation from \citet{Lyne2010}, but consistent with that inferred by \citet{Shaw2022} at the 68\% confidence interval.
The power-law amplitude by comparison is substantially lower, resulting from the larger number of pulsars with lower $\delta\dot{\nu}$ than those analysed by \citet{Lyne2010} and \citet{Shaw2022}.

We also tested whether $\delta\dot{\nu}$ preferentially scales according to various derived pulsar parameters, such as characteristic age ($\tau_{c}$), implied surface dipole magnetic field strength ($B_{\rm surf}$) and rate of rotational kinetic energy loss ($\dot{E}$).
This was achieved by including a dependence on $\nu$ in the right-hand side of the scaling relation in Equation~\ref{eqn:nudot_scale} as
\begin{equation}
    \delta\dot{\nu} = 10^{\xi}\, \nu^{a}\, |\dot{\nu}|^{b}.
\end{equation}
Using the same Gaussian likelihood function as before, we found the change in spin-down rate follows the scaling relation
\begin{equation}
    \delta\dot{\nu} = 10^{-4.1 \pm 0.7}\, \nu^{-0.18 \pm 0.17}\, |\dot{\nu}|^{0.88 \pm 0.05},
\end{equation}
which is close to being consistent with the same $\delta\dot{\nu} \propto |\dot{\nu}|$ scaling relation from Equation~\ref{eqn:f1_relation}. 
The posterior distribution for the $\nu$ scaling parameter overlaps with zero at the 95\% confidence interval, indicating there is little dependence between the amount of spin-down variability in a given pulsar and its spin frequency.


\subsection{Implications for precision timing experiments}

Millisecond pulsars are known to display emission state switching on short timescales \citep{Mahajan2018, Miles2022, Nathan2023} and transient profile shape changes \citep{Shannon2016, Lam2018, Goncharov2021, Jennings2024}, with the latter effect requiring either profile-domain timing techniques to correct for them or careful removal of the corresponding signal in the timing residuals (see, e.g. \citealt{Shannon2016}).
The propensity of spin-down rate variations among our pulsar sample and clear positive scaling in variability amplitude with $\dot{\nu}$ suggests that this phenomenon may also be both present and detectable among the millisecond pulsar population.
Accurately correcting for these effects is particularly important for precision timing experiments aiming to detect the stochastic gravitational-wave background.

Recent hints of such a gravitational-wave background signal among pulsar timing experiments have revealed an unexpectedly large amplitude given previous upper limits \citep{Agazie2023b, EPTA2023b, Xu2023, Reardon2023, Miles2025b}, which appears to be increasing with time \citep{Reardon2023}.
It is presently unclear whether this is a genuine feature of the common signal, in which case it would be inconsistent with expectations for an isotropic gravitational-wave background, but could be the result of an unmodelled process intrinsic to the observed millisecond pulsars.  
Spin-down rate changes similar to those among our P574 pulsars would manifest as a form of non-stationary timing noise, for which the red power-law models that are typically used to account for timing noise have been demonstrated as insufficient for fully characterising the resulting timing variations \citep{Keith2023, Lower2023}. 

Assuming the rotational characteristics of millisecond pulsars follow the same scaling as the young pulsar population, our results can be used to predict the expected spin-down rate variability among the pulsars currently monitored by pulsar timing arrays.
The spin-down rates of the pulsars included in the upcoming third data release of the International Pulsar Timing Array (K. Liu, private communication; \citealt{Zic2023, EPTA2023a, Agazie2023a, Miles2025a}) largely range between $\dot{\nu} = -10^{-16}$\,s$^{-2}$ and $-10^{-14}$\,s$^{-2}$, alongside several noteworthy outliers such as PSR~B1937$+$21 with its high spin-down rate of $-4.3 \times 10^{-14}$\,s$^{-2}$. 
If the variable spin-down rate process is active among them, then from Figure~\ref{fig:nudot_pop} we would expect these pulsars to display changes in $\dot{\nu}$ with amplitudes ranging between $\delta\dot{\nu} = 10^{-19}$ to $10^{-15}$\,s$^{-2}$. 
This would be easily discernable using the Gaussian process regression framework that we employed. 

\subsection{Quasi-periodic variability}

Among our broader sample there exists a sub-group of 45 pulsars that display striking quasi-periodic oscillations in their spin-down timeseries.
These pulsars were predominantly identified through a combination of visual inspection of their timing residuals, spin-down timeseries and Lomb-Scargle periodograms.
This group includes five that were previously identified by \citet{Lyne2010} and \citet{Kerr2016}.
Modulation periods ($\mathcal{P}$) listed in Table~\ref{tbl:qpo} were recovered from Lomb-Scargle periodograms generated from their spin-down timeseries.
We failed to recover the sub-annual periodicities of 120\,d and 123\,d in PSRs J1646$-$4346 and J1825$-$1446 that were found by \citet{Kerr2016}. This may be due to a combination of our extended timing baselines and differences in methodology.
Ten of the pulsars in Table~\ref{tbl:qpo} display both long and short modulation timescales that may be harmonically related.
The `long' modulation periods in PSRs J1136$-$5525, J1243$-$6423, J1649$-$4653, J1705$-$4306 and J1830$-$1059 are either close to or exactly twice the duration of the corresponding `short' periods in these pulsars.
Hence, the short periods may be second harmonics.
In comparison the dual modulation timescales in PSRs J1056$-$6258 J1224$-$6407, J1326$-$6700, J1600$-$5751 and J1722$-$3712 display approximate fractional relations of 5/2, 7/4, 9/5, 3/2 and 7/5 respectively.

\begin{table}
\begin{center}
\caption{List of pulsars with highly quasi-periodic spin-down variations and their associated modulation periods. \label{tbl:qpo}}
\begin{tabular}{lccc}
\hline
PSR & $\mathcal{P}_{\rm short}$ (d) & $\mathcal{P}_{\rm long}$ (d) & $\mathcal{P}_{\rm single}$ (d) \\
\hline
J0255$-$5304 (B0254$-$53)  & - & - & 1648 \\
J0614$+$2229 (B0611$+$22)  & - & - & 268 \\
J1056$-$6258 (B1054$-$62)  & 177 & 457 & - \\
J1136$-$5525 (B1133$-$55)  & 598 & 1141 & - \\
J1224$-$6407 (B1221$-$63)  & 427 & 747 & - \\
J1243$-$6423 (B1240$-$64)  & 626 & 1203 & - \\
J1306$-$6617 (B1303$-$66)  & - & - & 1522 \\
J1326$-$5859 (B1323$-$58)  & - & - & 1272 \\
J1326$-$6700 (B1322$-$66)  & 232 & 437 & - \\
J1327$-$6301 (B1323$-$627) & - & - & 712 \\
J1327$-$6400               & - & - & 1505 \\
J1352$-$6803               & - & - &  508 \\
J1418$-$3921               & - & - &  911 \\
J1428$-$5530 (B1424$-$55)  & - & - & 1395 \\
J1514$-$5925               & - & - & 258 \\
J1534$-$5405 (B1530$-$539) & - & - & 856 \\
J1548$-$5607               & - & - & 1056 \\
J1600$-$5044 (B1557$-$50)  & - & - & 1549 \\
J1600$-$5751 (B1556$-$57)  & 477 & 739 & - \\
J1601$-$5335               & - & - & 1098 \\
J1626$-$4807               & - & - & 277 \\
J1637$-$4642               & - & - & 187 \\
J1611$-$5209 (B1607$-$52)  & - & - & 1671 \\
J1638$-$4417               & - & - & 982 \\
J1638$-$4608               & - & - & 478 \\
J1649$-$4653               & 1214 & 2541 & - \\
J1702$-$4306               & - & - & 392 \\
J1705$-$3950               & 157 & 314 & - \\
J1716$-$4005               & - & - & 627 \\
J1717$-$3425 (B1714$-$34)  & - & - & 1422 \\
J1717$-$5800               & - & - & 562 \\
J1726$-$3530               & - & - & 1066 \\
J1722$-$3632               & - & - & 438 \\
J1722$-$3712               & 105 & 148 & - \\
J1739$-$3023               & - & - & 235 \\
J1801$-$2154               & - & - & 170 \\
J1807$-$0847 (B1804$-$08)  & - & - & 988 \\
J1816$-$2650 (B1813$-$26)  & - & - & 1765 \\
J1829$-$1751 (B1826$-$17)  & - & - & 807 \\
J1830$-$1059 (B1828$-$11)  & 221 & 456 & - \\
J1835$-$0944               & - & - & 524 \\
J1853$-$0004               & - & - & 2118 \\
J1853$+$0011               & - & - & 484 \\
J1941$-$2602 (B1937$-$26)  & - & - & 1098 \\
J2346$-$0609               & - & - & 966 \\
\hline
\end{tabular}
\end{center}
\end{table}

Quasi-periodicities in the timing of several pulsars have been known about for decades, with free-precession often being invoked as the `clock' responsible for driving the oscillations (e.g. \citealt{Stairs2000}).
Free precession has also been postulated as the driver of the helical structure in the X-ray jet from the Vela pulsar \citep{Durant2013} and periodic variations in the inferred magnetic inclination angle of a radio-loud magnetar \citep{Desvignes2024}.
The discovery of short-timescale mode switching in these variable pulsars has been argued as evidence against free-precession with the correlation between profile shape and spin-down changes resulting from pulsars preferentially spending a greater amount of time in one magnetospheric state over another \citep{Lyne2010, Stairs2019}.
However, \citet{Jones2012} pointed out that free-precession may act as a trigger for the delicately balanced pulsar magnetospheres to move between different emission states.
In doing so, the observational effects of precession are amplified.

\begin{figure}
    \centering
    \includegraphics[width=\linewidth]{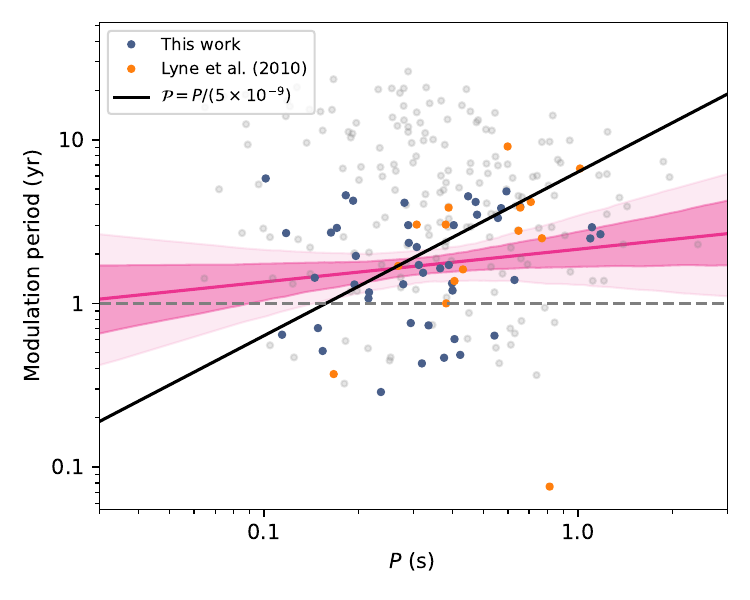}
    \caption{Comparison of spin-down modulation period with $P$. Grey points are the dominant quasi-periods of all pulsars indicated by their Lomb-Scargle periodograms. Dark blue points indicate the highly quasi-periodic pulsars listed in Table~\ref{tbl:qpo}, while the orange points are from Table 15 of \citet{Lyne2010}. The solid black curve is the $P/\mathcal{P} = 5\times 10^{-9}$ relation for freely precessing pulsars from \citet{Jones2012}, while the pink line and shading indicates our median fit along with the 68\% and 95\% confidence intervals. The dashed horizontal line indicates a quasi-period of 1\,yr.}
    \label{fig:qpo_p0}
\end{figure}

\begin{figure}
    \centering
    \includegraphics[width=\linewidth]{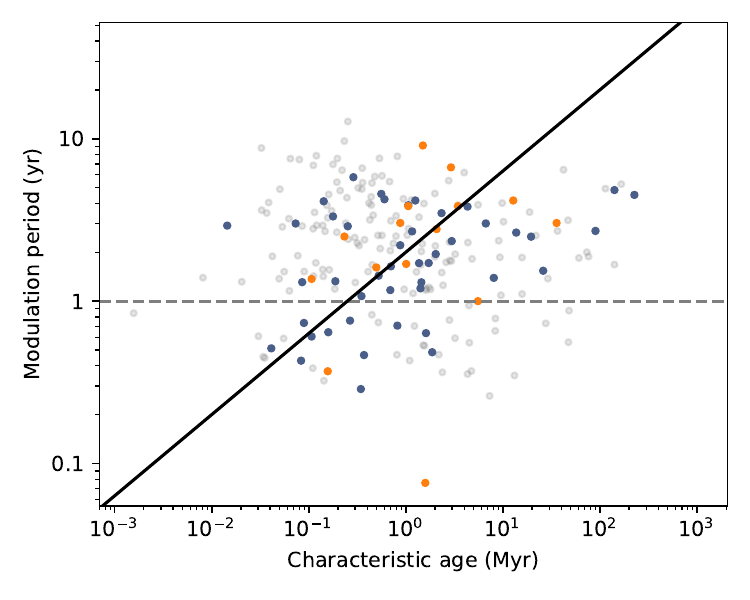}
    \caption{As Figure~\ref{fig:qpo_p0}, but for pulsar characteristic age ($\tau_{c}$). The solid black line indicates the relation from equation 42 of \citet{Jones2012} for $H_{c} = 10^{15}$\,G.}
    \label{fig:qpo_age}
\end{figure}

\begin{figure}
    \centering
    \includegraphics[width=\linewidth]{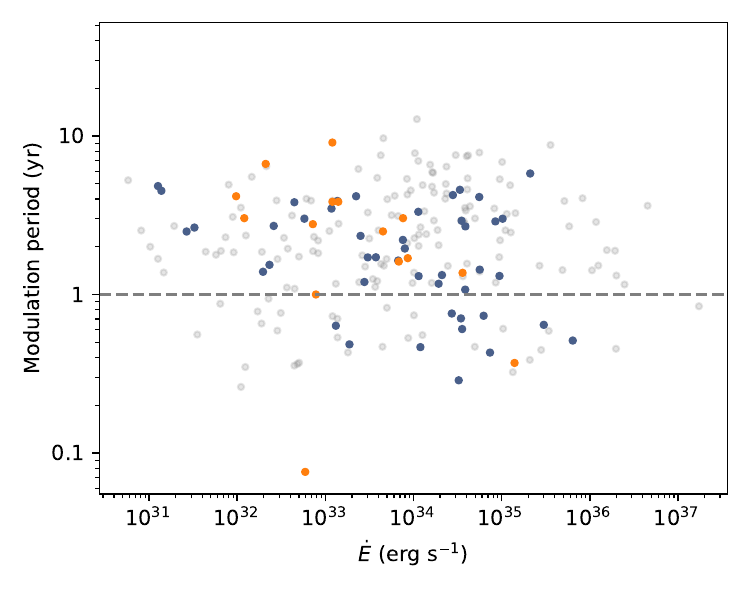}
    \caption{As Figure~\ref{fig:qpo_p0}, but for spin-down energy ($\dot{E}$).}
    \label{fig:qpo_edot}
\end{figure}

Several possible scaling relations between the free-precession induced modulation quasi-periods ($\mathcal{P}$) were derived by \citet{Jones2012} for various mechanisms that could result in deviations in the neutron star shape away from axisymmetry.
Relaxation of the crust or magnetic deformation were excluded as possible mechanisms on the basis that the resulting modulation periods were far in excess of those observed among the \cite{Lyne2010} sample.
Precession driven by elastic strain gives rise to a simple relation with spin period ($P$) of $P/\mathcal{P} = 5\times10^{-9}$, which both the \citet{Lyne2010} and an extended sample analysed by \citet{Kerr2016} largely follow.
Yet this relation does not appear to hold for our much larger quasi-periodic pulsar sample in Table~\ref{tbl:qpo}.
Our larger sample of modulation periods shown in Figure~\ref{fig:qpo_p0} does not display a strong dependence on the pulsar spin period. 
The amount of scatter becomes even larger when considering the apparent quasi-periods among the full sample of pulsars shown in grey.
The Spearman rank correlation coefficient of $r_{s} = 0.03$ (p-value $= 0.8$) indicates no correlation between these two values.
Performing a power-law fit to our sample of modulation periods and spin-periods returns the scaling relation
\begin{equation}
\mathcal{P} = 10^{0.3 \pm 0.1}\,{\rm yr}\, \Bigg( \frac{P}{1\,{\rm s}} \Bigg)^{0.2 \pm 0.2},
\end{equation}
which is inconsistent with the free-precession relation.
While it is consistent with the $\mathcal{P} \approx 1.4\,{\rm yr}\, (P/1\,{\rm s})^{1/2} (\lambda/10^{6}\,{\rm cm})$ scaling relation predicted for Tkachenko waves when assuming an oscillation wavelength of $2\times 10^{6}$\,cm at the 95\% confidence interval (see figure 5 in \citealt{Jones2012}), this is likely just a coincidence given the scatter among our measurements.
We find similar inconsistencies with the proposed scaling with $\tau_{c}$ for precession due to a superconducting core.
In Figure~\ref{fig:qpo_age} we show the highly scattered measurements do not follow the predicted relationship, with a Spearman coefficient of $r_{s} = 0.04$ (p-value $= 0.8$).
We also do not recover the relation with $\dot{E}$ tentatively identified by \citet{Kerr2016}, as shown in Figure~\ref{fig:qpo_edot} where $r_{s} = -0.02$ (p-value $= 0.9$).

The lack of consistency between our much larger sample of pulsars with quasi-period $\dot{\nu}$ variations and these predicted relations suggests that simple models of elastic strain or superconductivity driven free precession are not the dominant drivers of this behaviour.
\citet{Jones2012} noted that the relationships with $P$ required all pulsars display similar levels of strain in their crusts, while superconductivity necessitated that the entire star undergo precession.
Relaxing these assumptions would resolve these two issues, but would be difficult to falsify.
Another complicating factor is the presence of glitches among many of these pulsars, which can arise from interactions between the internal superfluid and the crust.
Pinning of superfluid vortices within the crust is expected to dampen free-precession on relatively short timescales \citep{Jones2001, Link2001, Jones2017}.
Non-radial oscillation modes of varying velocities and amplitudes have been suggested as an alternate means of driving changes in pulsar emission and spin-down, though without a clear mechanism \citep{Rosen2011}.
Hall waves launched by transient events both in the superfluid core and crust of neutron stars have been recently proposed as a means of generating long-term quasi-periodic oscillations in $\dot{\nu}$ that affect the inferred pulsar braking index \citep{Bransgrove2025}.
The periods of the waves investigated are significantly longer (1-100\,kyr) than the quasi-periodic processes we observe, but shorter period waves could be launched by starquakes acting as glitch triggers \citep{Bransgrove2020}.

\subsection{Transient events and interactions with planetesimals}

One mechanism that has been previously highlighted as a means for triggering both short-lived transient and persistent step-changes in $\dot{\nu}$ are interactions between a pulsar and infalling small objects, such as asteroids \citep{Brook2014, Brook2016, Lower2023}.
In this scenario, a minor body falls towards the pulsar and is vaporised by the extreme high-energy electromagnetic radiation environment and intense particle wind surrounding the neutron star.
The ionised remains of this object are then funnelled onto the magnetic poles of the pulsar, altering the plasma content which in-turn enhances/attenuates the existing radio emission, or activates new regions above the magnetic poles.
These modifications to the magnetospheric plasma will also alter the torque acting to slow the neutron star over time.
The altered emission and spin-down would then decay back to the original state as the excess of charged particles from the ionised body dissipates.
Assuming the infalling object is completely ionised, we can infer the resulting change in magnetospheric charged particle density from the step change in $\dot{\nu}$ via \citep{Kramer2006} 
\begin{equation}
    \Delta\rho = \frac{3I\delta\dot{\nu}}{R_{\rm pc}^{4} B_{0}},
\end{equation}
where $I = 10^{45}$\,g\,cm$^{-3}$ is the canonical neutron star moment of inertia, $R_{\rm pc} = \sqrt{2\pi R^{3} \nu/c}$ is the polar cap radius (we assume a standard neutron star radius of $R = 10^{6}$\,cm) and $B_{0} = 3.2\times 10^{19}\,{\rm G} \sqrt{-\dot{\nu}/\nu^{3}}$ is the surface magnetic field strength assuming the pulsar is spinning down purely via dipole radiation. 
The mass of the ionised minor body can then be determined from \citep{Brook2014}
\begin{equation}
    m = c\, \Delta\rho R_{\rm pc} \Delta t,
\end{equation}
where $c$ is the vacuum speed of light and $\Delta t$ is the duration in which the pulsar remains in the altered spin-down state.

Out of the 238 pulsars in Figure~\ref{fig:nudot} with variable spin-down rates, we were able to visually identify 68 individual transient spin-down events in 26 pulsars. 
The absolute size and duration of these $\dot{\nu}$ events along with the inferred asteroid masses are listed in Table~\ref{tbl:ast}.
We caution that this sample is incomplete to small transient changes in $\dot{\nu}$, which are inherently difficult to distinguish from other stochastic variations and sinusoidal changes in $\dot{\nu}$ arising from residual positional offsets.
Comparing the putative asteroid masses, the smallest resulted in the second event in PSR~J1833$-$0827 with $m \sim 2.6 \times 10^{11}$\,g. 
The third (MJD 54600-55400) event in PSR~J1602$-$5100 is the largest with a mass of $5.32 \times 10^{14}$\,g.
This is however smaller than the $\sim 3\times10^{15}$\,g mass inferred for the large 2005 spin-down/profile event detected in PSR~J0738$-$4042 \citep{Brook2014, Lower2023}.
Under the simplifying assumption that these objects were approximately spherical and C-type (carbonaceous) asteroids with a density of $\sim 1.57$\,g\,cm$^{-3}$ \citep{Carry2012}, the computed diameters range between $6.8 \times 10^{3}$ to $7.5 \times 10^{4}$\,cm and $1.5 \times 10^{6}$\,cm for the 2005 event in PSR~J0738$-$4042.
Objects within this size range are exceedingly common among main-belt asteroids in the Solar System \citep{Durda1998}.
If these transient spin-down events are indeed the result of pulsar-asteroid interactions, then their prevalence implies that debris discs and asteroids orbiting pulsars should be relatively common.
Previous searches for planetesimals orbiting pulsars have largely been unsuccessful, though the lowest accessible mass ranges of these searches have been predominantly limited by the achieved timing precision. 
Surveys of hundreds of slow pulsars placed stringent limits on the presence of orbiting bodies larger than the mass of Mercury \citep{Kerr2015, Nitu2022}, while analyses of millisecond pulsar timing arrays have pushed these limits to between $\sim 0.001$-$1$\,Lunar masses \citep{Behrens2020, Nitu2024}.
These limits are several orders of magnitude higher than the putative asteroid masses listed in Table~\ref{tbl:ast}.

Exactly where these asteroids would have originated remains an open question.
Fall-back of supernova material and relic discs from the progenitor stellar system are both popular mechanisms for producing asteroid belts around pulsars \citep{Michel1988}.
However the high kick velocities measured across the pulsar population present a challenge to these scenarios \citep{Hobbs2005}.
Stripping of material from a (former) companion star by a kicked neutron star is one means of forming a disc around high-velocity pulsars, though the expected rate of such interactions is vanishingly small for natal kicks below $\lesssim 600$\,km\,s$^{-1}$ (see Figure 10 in \citealt{Hirai2022}).
The longevity of debris discs around pulsars and condensation rate of objects is also subject to other factors, such as a reduction in pulsar wind energy due to pulsar spin-down evolution \citep{Menou2001}.

Interactions with free-floating interstellar objects (ISOs) similar to 1I/`Oumuamua and 2I/Borisov may present another possible origin for these events.
Direct impacts between minor bodies and neutron stars have been proposed as a potential mechanism for producing both gamma-ray bursts (GRBs; \citealt{Newman1980}) and fast radio bursts (FRBs; \citealt{Geng2015, Dai2016}).
The ISO/neutron star collision rate of $\sim 10^{7}$\,Gpc$^{-3}$\,yr$^{-1}$ (or $\sim 10^{-7}$\,yr$^{-1}$ for an individual neutron star) derived by \citet{Pham2024} is consistent with the observed all-sky FRB rate of $(7^{+9}_{-6}) \times 10^{7}$\,Gpc$^{-3}$\,yr$^{-1}$ \citep{Bochenek2020}.
Such collisions would avoid the challenges associated with forming or retaining asteroid belts around high-velocity pulsars. 
To test this scenario we computed the rate of observed transient spin-down events occur in our pulsar sample. 
For $n_{|\delta\dot{\nu}|} = 70$ events observed in $f_{|\delta\dot{\nu}|} = 27/260$ pulsars (note, this includes the two events in PSR~J0738$-$4042) over a total accumulated monitoring time of $T_{\rm acc} = 4095.6$\,yr, the rate can be derived as
\begin{equation}
    N_{|\delta\dot{\nu}|} \propto (n_{|\delta\dot{\nu}|}/T_{\rm acc}) \times f_{|\delta\dot{\nu}|} \sim 0.002\,{\rm yr}^{-1}.
\end{equation}
This is over four orders of magnitude higher than the per-pulsar rate from \citet{Pham2024}.
While their rate is an underestimate due to not taking into account interactions between infalling ISOs and pulsar radiation beams or evaporation from high-energy flux and particle winds, it would likely remain difficult to fully reconcile the difference.
Many of these pulsars display repeat transient spin-down events which is also inconsistent with the implied per-pulsar ISO collision rate.
Other tests, such as the detection of FRBs from our pulsar sample, are impractical given the short duration of our Parkes observations and monthly cadence. 
There are however no reports of GRBs detected by all-sky monitors as having originated from any of our pulsars.

Gravitational scattering of ISOs on hyperbolic encounters can produce cuspy signals in pulsar timing residuals, where the sharpness and sign of the cusp depends on the viewing geometry and periastron distance, and the amplitude on the ISO mass \citep{Jennings2020a}.
Close encounters of around 1\,AU can produce signals that can appear indistinguishable from a glitch given a low observation cadence. 
More distant interactions would result in a more rounded peak, strikingly similar to those associated with the timing events in several pulsars displayed in Figure~\ref{fig:resids}.
However, given the linear scaling between the timing perturbation and ISO mass (Equation 16 in \citealt{Jennings2020a}), we can infer that interactions ISOs with masses equivalent to that of gas giant planets (i.e $\sim 10^{-4}$\,M$_{\odot}$) would be required to reproduce the peaks of between 10's-100's of milliseconds seen among our pulsar sample.
It is also unclear how such distant encounters would alter the radio emission mechanism of the pulsar.

\section{Conclusions}\label{sec:conc}

We have used a combination of Gaussian process regression and Bayesian inference techniques to demonstrate that variations in profile shape and spin-down rate are pervasive among the population of radio pulsars.
Our sample of 238 pulsars with significant fluctuations in spin-down and 52 pulsars that display profile shape changes is the largest assembled to date.
These include 29 pulsars for which we describe the links between their variable emission and spin-down for the first time.
Using the inferred min/max differences in spin-down rate timeseries, we demonstrated the $\dot{\nu}$ variability amplitude scales with spin-down rate according to $\delta\dot{\nu} \propto 10^{-4.5} |\dot{\nu}_{\rm weak}|^{0.85}$. 
No substantial dependence on spin-frequency was identified, which could have strong implications for the prevalence of these behaviours in millisecond pulsars if they follow the a similar scaling relation with spin-down rate.
Manifestation of spin-down variations in millisecond pulsar timing arrays could leak into the detected common signal seen, as the resulting non-stationary noise process are not well characterised by standard power-law red noise models.
Properly accounting for this effect may both improve gravitational-wave searches, and resolve challenges surrounding the apparent time-dependence of the putative gravitational-wave signal, motivating the application of the spin-down modelling to determine whether this effect is present among these pulsars.

Alongside this, we showed that the various expected scaling relations from models of free-precession, a popular mechanism for producing highly quasi-periodic changes in profile shape and spin-down, are unable to fully account for the large spread in modulation periods seen among a sub-group of 45 pulsars that display extremely (quasi-)periodic spin-down rate (and occasionally, profile shape) variations.
This lends weight towards a magnetospheric state-switching origin for the quasi-periodic effects. 
We also re-examined interactions between pulsars and infalling planetesimals as a means of producing transient changes in profile and spin-down rate among 26 pulsars.
Direct collisions and hyperbolic flybys of interstellar objects are ruled out, with the rate of transient spin-down events among these pulsars being four orders of magnitude higher than the theoretical impact rate, whereas repeat flybys of Jupiter-mass objects are required to explain the multiple large transient events seen in individual pulsars.
Disruption of infalling asteroids from debris disks orbiting these pulsars remains a valid explanation.
Assuming C-type composition asteroids, the masses and radii inferred from the amplitude and duration of these events are consistent with planetesimals that are prevalent in our Solar System.
Future direct detections of thermal emission from asteroid belts and debris disks around these pulsars via deep infrared or sub-mm observations would lend substantial weight towards this hypothesis.
A purely magnetospheric origin for these behaviours cannot be ruled out at present.

An apparent deficit in the number of pulsars displaying profile shape changes, despite showing clear $\dot{\nu}$ variations, is largely due to a combination of relatively low per-epoch signal-to-noise ratios, short observations resulting in stochastic jitter-dominated shape changes and smearing of profile components by scatter-broadening in the interstellar medium.
Similar challenges to identifying profile-shape changing pulsars have been faced by other large-scale timing programmes. 
This includes the Thousand Pulsar Array on MeerKAT, where few new shape-changing pulsars have been identified to date despite the substantially higher telescope gain over Murriyang \citep{Basu2024}. 
Increasing this population could be achieved through a re-optimisation of existing programmes and those proposed to take place using the SKA.
Lengthening the duration of each pulsar observation would reduce the impact of pulse-to-pulse jitter, and make such programmes sensitive to more subtle shape changes. 

\section*{Acknowledgements}

Murriyang, the Parkes radio telescope, is part of the Australia Telescope National Facility (\href{https://ror.org/05qajvd42}{https://ror.org/05qajvd42}) which is funded by the Australian Government for operation as a National Facility managed by CSIRO.
We acknowledge the Wiradjuri people as the traditional owners of the Observatory site.
This project was supported by resources and expertise provided by CSIRO IMT Scientific Computing.
Some of the data analysis behind this work was performed using the OzSTAR national HPC facility, which is funded by Swinburne University of Technology and the National Collaborative Research Infrastructure Strategy (NCRIS). 
MEL is supported by an Australian Research Council (ARC) Discovery Early Career Research Award DE250100508.
PRB is supported by the UK Science and Technology Facilities Council (STFC), grant number ST/W000946/1.
RMS acknowledges support through ARC Future Fellowship FT190100155. 
Part of the work was undertaken with support from the ARC Centre of Excellence for Gravitational Wave Discovery (OzGrav; CE170100004 and CE230100016).
Work at NRL is supported by NASA.
Pulsar research at the Jodrell Bank Centre for Astrophysics is supported by a consolidated grant from the UK STFC.
This work has made use of NASA's Astrophysics Data System, and the following open-source software packages that were not cited elsewhere: {\sc matplotlib} \citep{matplotlib}, {\sc numpy} \citep{numpy} and {\sc scipy} \citep{scipy}.
MEL is grateful to the hospitality of St Edmund Hall during visits to Oxford throughout this project. 

\section*{Data Availability}

The raw data is available to download via the CSIRO Data Access Portal (\href{https://data.csiro.au/}{https://data.csiro.au/}) following an 18-month proprietary period. 
Other data products are available upon reasonable request to the corresponding author.



\bibliographystyle{mnras}
\bibliography{main} 



\appendix

\section{Spin-down model parameters and inferred magnetospheric properties}\label{appndx}

Tables listing the best fit model parameters for the timing residual Gaussian process regression fits, pulsar rotational and derived magnetospheric parameters, and transient spin-down event/asteroid properties.

\begin{table*}
\begin{center}
\caption{Results of our Gaussian process fits, including preferred model, Bayesian information criterion, maximum likelihood hyperparameters and corresponding $\mathcal{K}$-metric. The $\sigma_{2}^{2}$ and $\lambda_{2}$ parameters are left blank for pulsars where only a single kernel was preferred. \label{tbl:nudot_fit}}
\begin{tabular}{lccccccccc}
\hline
PSR & MJD range & $N_{\rm ToA}$ & $\Delta{\rm BIC}$ & $\sigma_{1}^{2}$ (s) & $\lambda_{1}$ (d) & $\sigma_{2}^{2}$ (s) & $\lambda_{2}$ (d) & $\sigma_{N}^{2}$ (s) & $\ln\mathcal{K}$ \\
\hline
J0034$-$0721 & 53040-60102 & 164 & $10.2$ & $8.9\times 10^{-7}$ & $915.8$ & - & - & $1.2\times 10^{-5}$ & $9.5$ \\
J0108$-$1431 & 49091-60037 & 312 & $5.5$ & $9.4\times 10^{-7}$ & $857.3$ & - & - & $1.1\times 10^{-6}$ & $8.7$ \\
J0134$-$2937 & 53972-60101 & 136 & $7.3$ & $6.5\times 10^{-8}$ & $1787.8$ & - & - & $1.9\times 10^{-12}$ & $18.4$ \\
J0152$-$1637 & 51874-60102 & 134 & $9.7$ & $3.4\times 10^{-6}$ & $2304.5$ & - & - & $2.4\times 10^{-8}$ & $13.9$ \\
J0206$-$4028 & 51874-60102 & 134 & $-22.7$ & $5.3\times 10^{-3}$ & $1945.1$ & $1.0\times 10^{-7}$ & $63.4$ & $4.6\times 10^{-8}$ & $7.2$ \\
J0255$-$5304 & 51530-60100 & 539 & $11.8$ & $2.1\times 10^{-9}$ & $476.8$ & - & - & $4.2\times 10^{-8}$ & $10.9$ \\
J0401$-$7608 & 54220-60100 & 230 & $-16.7$ & $7.7\times 10^{-3}$ & $1156.4$ & $4.9\times 10^{-8}$ & $96.1$ & $8.6\times 10^{-8}$ & $9.1$ \\
\hline
\end{tabular}
\end{center}
\end{table*}
\begin{table*}
\begin{center}
\contcaption{}
\begin{tabular}{lccccccccc}
\hline
PSR & MJD range & $N_{\rm ToA}$ & $\Delta{\rm BIC}$ & $\sigma_{1}^{2}$ (s) & $\lambda_{1}$ (d) & $\sigma_{2}^{2}$ (s) & $\lambda_{2}$ (d) & $\sigma_{N}^{2}$ (s) & $\ln\mathcal{K}$ \\
\hline
J1157$-$6224 & 56672-60100 & 135 & $8.0$ & $6.2\times 10^{-7}$ & $191.1$ & - & - & $5.0\times 10^{-8}$ & $11.3$ \\
J1216$-$6223 & 50852-56708 & 108 & $8.1$ & $9.6\times 10^{-4}$ & $1112.4$ & - & - & $5.4\times 10^{-7}$ & $14.3$ \\
J1224$-$6407 & 48361-60137 & 700 & $-227.8$ & $1.5\times 10^{-2}$ & $3000.0$ & $9.3\times 10^{-7}$ & $169.7$ & $1.9\times 10^{-9}$ & $14.6$ \\
J1243$-$6423 & 56672-60100 & 129 & $-17.9$ & $3.3\times 10^{-7}$ & $131.5$ & $1.0\times 10^{-4}$ & $507.9$ & $4.3\times 10^{-9}$ & $11.6$ \\
J1248$-$6344 & 51215-56709 & 123 & $-8.6$ & $1.3\times 10^{-6}$ & $130.4$ & $1.8\times 10^{-3}$ & $1021.4$ & $6.7\times 10^{-7}$ & $12.4$ \\
J1305$-$6203 & 50941-60137 & 208 & $-1.9$ & $2.9\times 10^{-5}$ & $751.7$ & $1.2\times 10^{-1}$ & $3000.0$ & $4.5\times 10^{-7}$ & $15.1$ \\
J1306$-$6617 & 56672-60100 & 109 & $-7.8$ & $1.1\times 10^{-3}$ & $669.0$ & $6.4\times 10^{-7}$ & $157.9$ & $7.0\times 10^{-8}$ & $10.7$ \\
\hline
\end{tabular}
\end{center}
\end{table*}
\begin{table*}
\begin{center}
\contcaption{}
\begin{tabular}{lccccccccc}
\hline
PSR & MJD range & $N_{\rm ToA}$ & $\Delta{\rm BIC}$ & $\sigma_{1}^{2}$ (s) & $\lambda_{1}$ (d) & $\sigma_{2}^{2}$ (s) & $\lambda_{2}$ (d) & $\sigma_{N}^{2}$ (s) & $\ln\mathcal{K}$ \\
\hline
J1623$-$4256 & 56682-60079 & 106 & $-8.1$ & $3.1\times 10^{-6}$ & $191.6$ & $1.7\times 10^{-3}$ & $833.0$ & $3.1\times 10^{-8}$ & $9.8$ \\
J1626$-$4807 & 50941-57840 & 135 & $-49.8$ & $2.2\times 10^{-5}$ & $75.0$ & $2.1\times 10^{-2}$ & $1064.1$ & $6.3\times 10^{-6}$ & $9.2$ \\
J1627$-$4706 & 52807-56709 & 121 & $9.6$ & $1.3\times 10^{-6}$ & $183.9$ & - & - & $7.4\times 10^{-7}$ & $11.5$ \\
J1630$-$4733 & 57165-60079 & 86 & $5.8$ & $1.6\times 10^{-3}$ & $243.7$ & - & - & $1.5\times 10^{-6}$ & $10.1$ \\
J1632$-$4621 & 56682-60079 & 101 & $-12.4$ & $2.5\times 10^{-3}$ & $876.0$ & $4.2\times 10^{-7}$ & $129.5$ & $1.0\times 10^{-7}$ & $10.2$ \\
J1632$-$4757 & 51467-56709 & 116 & $0.9$ & $9.0\times 10^{-3}$ & $697.4$ & - & - & $1.8\times 10^{-6}$ & $13.6$ \\
J1632$-$4818 & 50853-58500 & 206 & $-97.3$ & $1.0e+01$ & $1384.2$ & $1.3\times 10^{-4}$ & $145.1$ & $7.2\times 10^{-6}$ & $11.9$ \\
\hline
\end{tabular}
\end{center}
\end{table*}
\begin{table*}
\begin{center}
\contcaption{}
\begin{tabular}{lccccccccc}
\hline
PSR & MJD range & $N_{\rm ToA}$ & $\Delta{\rm BIC}$ & $\sigma_{1}^{2}$ (s) & $\lambda_{1}$ (d) & $\sigma_{2}^{2}$ (s) & $\lambda_{2}$ (d) & $\sigma_{N}^{2}$ (s) & $\ln\mathcal{K}$ \\
\hline
J1743$-$3150 & 56682-59405 & 74 & $7.6$ & $4.3\times 10^{-6}$ & $307.2$ & - & - & $8.2\times 10^{-7}$ & $10.6$ \\
J1745$-$3040 & 51901-60138 & 299 & $-21.4$ & $8.8\times 10^{-2}$ & $2443.8$ & $1.7\times 10^{-7}$ & $374.5$ & $5.6\times 10^{-8}$ & $14.7$ \\
J1749$-$3002 & 56741-60079 & 97 & $0.6$ & $1.2\times 10^{-5}$ & $124.2$ & - & - & $2.2\times 10^{-7}$ & $8.9$ \\
J1750$-$3157 & 57166-60163 & 71 & $8.5$ & $6.2\times 10^{-9}$ & $210.0$ & - & - & $4.5\times 10^{-8}$ & $7.9$ \\
J1751$-$4657 & 56682-60101 & 97 & $9.1$ & $9.6\times 10^{-7}$ & $524.4$ & - & - & $1.5\times 10^{-8}$ & $11.3$ \\
J1752$-$2806 & 56741-60101 & 107 & $-12.8$ & $6.5\times 10^{-7}$ & $62.0$ & $1.8\times 10^{-4}$ & $357.0$ & $5.9\times 10^{-8}$ & $8.6$ \\
J1757$-$2421 & 51529-60162 & 251 & $-239.7$ & $3.1\times 10^{-7}$ & $101.9$ & $3.9\times 10^{-3}$ & $1064.8$ & $1.3\times 10^{-8}$ & $13.0$ \\
\hline
\end{tabular}
\end{center}
\end{table*}

\begin{table*}
\begin{center}
\caption{Measured pulsar rotational parameters from timing and spin-down modelling, alongside the derived magnetospheric properties. \label{tbl:magnetosphere}}
\begin{tabular}{lcccccc}
\hline
PSR & $\nu$ (Hz) & $\dot{\nu}_{\rm weak}$ (s$^{-2}$) & $\dot{\nu}_{\rm strong}$ (s$^{-2}$) & $\delta\dot{\nu}$ (s$^{-2}$) & $\delta\rho$ (mC\,m$^{-3}$) & $\delta\rho/\rho_{\rm GJ}$ (\%) \\
\hline
J0034$-$0721 & $1.0605$ & $-4.5871\times 10^{-16}$ & $-4.5922\times 10^{-16}$ & $5.0505\times 10^{19}$ & $0.016$ & $0.22$ \\
J0108$-$1431 & $1.2383$ & $-1.1772\times 10^{-16}$ & $-1.1874\times 10^{-16}$ & $8.5542\times 10^{19}$ & $0.05$ & $1.45$ \\
J0134$-$2937 & $7.3013$ & $-4.1775\times 10^{-15}$ & $-4.178\times 10^{15}$ & $3.9083\times 10^{19}$ & $0.002$ & $0.02$ \\
J0152$-$1637 & $1.2009$ & $-1.8745\times 10^{-15}$ & $-1.8748\times 10^{-15}$ & $3.3516\times 10^{19}$ & $0.005$ & $0.04$ \\
J0206$-$4028 & $1.5859$ & $-2.9555\times 10^{-15}$ & $-3.0835\times 10^{-15}$ & $1.28\times 10^{-16}$ & $1.331$ & $8.67$ \\
J0255$-$5304 & $2.2336$ & $-1.557\times 10^{16}$ & $-1.5597\times 10^{-16}$ & $2.6776\times 10^{19}$ & $0.01$ & $0.34$ \\
J0401$-$7608 & $1.834$ & $-5.1618\times 10^{-15}$ & $-5.2151\times 10^{-15}$ & $5.332\times 10^{-17}$ & $0.39$ & $2.07$ \\
\hline
\end{tabular}
\end{center}
\end{table*}

\begin{table*}
\begin{center}
\contcaption{}
\begin{tabular}{lcccccc}
\hline
PSR & $\nu$ (Hz) & $\dot{\nu}_{\rm weak}$ (s$^{-2}$) & $\dot{\nu}_{\rm strong}$ (s$^{-2}$) & $\delta\dot{\nu}$ (s$^{-2}$) & $\delta\rho$ (mC\,m$^{-3}$) & $\delta\rho/\rho_{\rm GJ}$ (\%) \\
\hline
J1157$-$6224 & $2.4967$ & $-2.448\times 10^{14}$ & $-2.453\times 10^{14}$ & $4.9665\times 10^{17}$ & $0.143$ & $0.41$ \\
J1216$-$6223 & $2.6734$ & $-1.2023\times 10^{-13}$ & $-1.2025\times 10^{-13}$ & $1.4214\times 10^{17}$ & $0.018$ & $0.02$ \\
J1224$-$6407 & $4.6194$ & $-1.0564\times 10^{-13}$ & $-1.0576\times 10^{-13}$ & $1.1901\times 10^{16}$ & $0.121$ & $0.23$ \\
J1243$-$6423 & $2.5741$ & $-2.9739\times 10^{-14}$ & $-2.9853\times 10^{-14}$ & $1.049\times 10^{-16}$ & $0.27$ & $0.71$ \\
J1248$-$6344 & $5.042$ & $-4.2978\times 10^{-13}$ & $-4.3015\times 10^{-13}$ & $3.6747\times 10^{16}$ & $0.178$ & $0.17$ \\
J1305$-$6203 & $2.3377$ & $-1.7571\times 10^{-13}$ & $-1.7576\times 10^{-13}$ & $5.6284\times 10^{17}$ & $0.063$ & $0.06$ \\
J1306$-$6617 & $2.114$ & $-2.6671\times 10^{-14}$ & $-2.6813\times 10^{-14}$ & $1.4216\times 10^{16}$ & $0.426$ & $1.07$ \\
\hline
\end{tabular}
\end{center}
\end{table*}

\begin{table*}
\begin{center}
\contcaption{}
\begin{tabular}{lcccccc}
\hline
PSR & $\nu$ (Hz) & $\dot{\nu}_{\rm weak}$ (s$^{-2}$) & $\dot{\nu}_{\rm strong}$ (s$^{-2}$) & $\delta\dot{\nu}$ (s$^{-2}$) & $\delta\rho$ (mC\,m$^{-3}$) & $\delta\rho/\rho_{\rm GJ}$ (\%) \\
\hline
J1623$-$4256 & $2.7428$ & $-7.5156\times 10^{-15}$ & $-7.6712\times 10^{-15}$ & $1.4618\times 10^{16}$ & $0.725$ & $3.89$ \\
J1626$-$4807 & $3.4022$ & $-2.0101\times 10^{-13}$ & $-2.0352\times 10^{-13}$ & $2.0203\times 10^{15}$ & $1.739$ & $2.01$ \\
J1627$-$4706 & $7.105$ & $-8.7275\times 10^{-14}$ & $-8.7434\times 10^{-14}$ & $1.5946\times 10^{16}$ & $0.144$ & $0.37$ \\
J1630$-$4733 & $1.7362$ & $-6.7028\times 10^{-14}$ & $-6.7759\times 10^{-14}$ & $7.3111\times 10^{16}$ & $1.526$ & $2.18$ \\
J1632$-$4621 & $0.5851$ & $-2.5973\times 10^{-14}$ & $-2.6009\times 10^{-14}$ & $3.2841\times 10^{17}$ & $0.19$ & $0.25$ \\
J1632$-$4757 & $4.3751$ & $-2.8825\times 10^{-13}$ & $-2.8868\times 10^{-13}$ & $4.3366\times 10^{16}$ & $0.275$ & $0.3$ \\
J1632$-$4818 & $1.229$ & $-9.7969\times 10^{-13}$ & $-9.8162\times 10^{-13}$ & $1.7697\times 10^{15}$ & $1.148$ & $0.36$ \\
\hline
\end{tabular}
\end{center}
\end{table*}

\begin{table*}
\begin{center}
\contcaption{}
\begin{tabular}{lcccccc}
\hline
PSR & $\nu$ (Hz) & $\dot{\nu}_{\rm weak}$ (s$^{-2}$) & $\dot{\nu}_{\rm strong}$ (s$^{-2}$) & $\delta\dot{\nu}$ (s$^{-2}$) & $\delta\rho$ (mC\,m$^{-3}$) & $\delta\rho/\rho_{\rm GJ}$ (\%) \\
\hline
J1745$-$3040 & $2.7216$ & $-7.9006\times 10^{-14}$ & $-7.9055\times 10^{-14}$ & $4.7583\times 10^{17}$ & $0.073$ & $0.12$ \\
J1749$-$3002 & $1.6397$ & $-2.1035\times 10^{-14}$ & $-2.1327\times 10^{-14}$ & $2.9123\times 10^{16}$ & $1.116$ & $2.77$ \\
J1750$-$3157 & $1.0985$ & $-2.365\times 10^{16}$ & $-2.3735\times 10^{-16}$ & $8.4797\times 10^{19}$ & $0.037$ & $0.72$ \\
J1751$-$4657 & $1.3471$ & $-2.3532\times 10^{-15}$ & $-2.3561\times 10^{-15}$ & $2.9172\times 10^{18}$ & $0.037$ & $0.25$ \\
J1752$-$2806 & $1.7776$ & $-2.55\times 10^{4}$ & $-2.5828\times 10^{-14}$ & $3.2806\times 10^{16}$ & $1.097$ & $2.57$ \\
J1757$-$2421 & $4.2716$ & $-2.355\times 10^{13}$ & $-2.3582\times 10^{-13}$ & $3.1374\times 10^{16}$ & $0.223$ & $0.27$ \\
J1801$-$2154 & $2.6645$ & $-1.1279\times 10^{-13}$ & $-1.1432\times 10^{-13}$ & $1.4381\times 10^{15}$ & $1.868$ & $2.55$ \\
\hline
\end{tabular}
\end{center}
\end{table*}

\begin{table}
\begin{center}
\caption{Transient spin-down event properties and inferred asteroid masses. \label{tbl:ast}}
\begin{tabular}{lccc}
\hline
PSR & $|\delta\dot{\nu}|$ ($10^{-16}$ s$^{-2}$) & $\Delta t$ (d) & $m$ ($10^{12}$ g) \\
\hline
J0627+0706 & 2.877 & 4153 & 87.92 \\
- & 2.575 & 889 & 16.85 \\
J0729-1836 & 3.166 & 1462 & 45.79 \\
- & 2.769 & 296 & 8.11 \\
J0809-4753 & 0.233 & 1385 & 8.49 \\
- & 0.108 & 524 & 1.49 \\
- & 0.174 & 518 & 2.37 \\
J0820-1350 & 0.02 & 631 & 0.89 \\
J0842-4851 & 0.499 & 331 & 2.91 \\
- & 0.676 & 643 & 7.65 \\
- & 1.232 & 516 & 11.19 \\
J1123-6259 & 4.01 & 569 & 22.82 \\
J1156-5707 & 25.04 & 2348 & 277.85 \\
J1340-6456 & 0.608 & 226 & 1.95 \\
- & 1.739 & 301 & 7.44 \\
- & 1.814 & 304 & 7.84 \\
J1401-6357 & 2.16 & 636 & 23.83 \\
- & 4.08 & 867 & 61.36 \\
J1412-6145 & 3.3 & 1159 & 10.23 \\
- & 7.17 & 3385 & 64.91 \\
J1539-5626 & 4.26 & 527 & 20.95 \\
- & 3.16 & 536 & 15.81 \\
- & 4.06 & 446 & 16.9 \\
- & 2.958 & 161 & 4.44 \\
- & 4.68 & 939 & 41.01 \\
- & 3.19 & 411 & 12.24 \\
J1541-5535 & 57.3 & 509 & 84.1 \\
- & 55.2 & 1059 & 168.57 \\
- & 26.2 & 502 & 37.93 \\
- & 26.1 & 311 & 23.41 \\
J1602-5100 & 10.7 & 625 & 58.57 \\
- & 13.0 & 624 & 71.04 \\
- & 42.0 & 958 & 352.39 \\
- & 11.8 & 538 & 55.6 \\
J1604-4909 & 1.447 & 1351 & 53.54 \\
- & 1.799 & 968 & 47.69 \\
J1632-4818 & 9.19 & 297 & 7.36 \\
J1733-3716 & 0.137 & 488 & 0.49 \\
- & 0.191 & 716 & 1.0 \\
- & 0.344 & 692 & 1.75 \\
- & 0.216 & 1497 & 2.38 \\
- & 0.183 & 1619 & 2.18 \\
J1734-3333 & 18.37 & 275 & 10.45 \\
- & 11.01 & 185 & 4.21 \\
J1806-2125 & 16.2 & 638 & 38.71 \\
- & 8.9 & 305 & 10.17 \\
- & 9.9 & 585 & 21.69 \\
- & 46.6 & 1732 & 302.26 \\
J1820-0427 & 0.962 & 959 & 18.52 \\
J1820-1529 & 12.04 & 508 & 27.95 \\
- & 17.88 & 556 & 45.43 \\
J1825-0935 & 20.44 & 426 & 78.13 \\
- & 6.04 & 309 & 16.75 \\
- & 7.3 & 268 & 17.55 \\
- & 9.62 & 285 & 24.6 \\
J1833-0827 & 0.419 & 460 & 0.46 \\
- & 0.238 & 459 & 0.26 \\
- & 0.489 & 544 & 0.63 \\
- & 0.43 & 834 & 0.85 \\
J1835-0643 & 6.88 & 4501 & 125.71 \\
J1838-0453 & 46.7 & 469 & 65.3 \\
\hline
\end{tabular}
\end{center}
\end{table}

\begin{table}
\begin{center}
\contcaption{}
\begin{tabular}{lccc}
\hline
PSR & $\delta\dot{\nu}$ ($10^{-16}$ s$^{-2}$) & $\Delta t$ (d) & $m$ ($10^{12}$ g) \\
\hline
J1841-0345 & 4.61 & 2858 & 29.84 \\
J1845-0434 & 4.33 & 879 & 46.47 \\
- & 2.14 & 542 & 14.16 \\
- & 2.12 & 911 & 23.58 \\
- & 0.88 & 296 & 3.18 \\
- & 4.98 & 1002 & 60.93 \\
- & 2.4 & 648 & 18.99 \\
\hline
\end{tabular}
\end{center}
\end{table}


\bsp	
\label{lastpage}
\end{document}